\documentclass[aps, onecolumn, pre, notitlepage,superscriptaddress, longbibliography, notitlepage]{revtex4-2}
\usepackage{bm,amsmath,graphicx, mathtools,multirow}
\usepackage{textcomp, gensymb}
\usepackage{graphicx}
\usepackage[colorlinks=true,linkcolor=MidnightBlue,urlcolor=black,citecolor=MidnightBlue,anchorcolor=MidnightBlue]{hyperref}
\usepackage[dvipsnames]{xcolor}
\usepackage{amsthm}

\usepackage{caption}
\begin{document}

\title{Mechanics and statistics of a solvable model of an autophoretic
colloidal chain}
\author{Arvin Gopal Subramaniam}
\thanks{ph22d800@smail.iitm.ac.in}
\affiliation{Department of Physics, Indian Institute of Technology Madras, Chennai, India}
\affiliation{Center for Soft and Biological Matter, IIT Madras, Chennai, India}

\author{Rajesh Singh}
\thanks{rsingh@physics.iitm.ac.in}
\affiliation{Department of Physics, Indian Institute of Technology Madras, Chennai, India}
\affiliation{Center for Soft and Biological Matter, IIT Madras, Chennai, India}

\begin{abstract}
Equilibrium statistical mechanics owes much of its analytical tractability to symmetry: detailed balance, gradient dynamics, and the resulting vanishing of steady-state entropy production are properties that follow directly from the structure of the underlying dynamics, not from any smallness of the perturbation. 
Exact solutions of this kind are rare away from equilibrium. 
Here we identify and characterize a class of far-from-equilibrium 
colloidal chains -- coupled through roto-translational and autophoretic interactions -- 
that nevertheless admit an exact \emph{quasi-equilibrium} description. 
For fixed chain geometries, the monomer orientational dynamics map directly onto a scalar potential. 
Detailed balance holds strictly within the orientational sector, causing its steady-state entropy production rate (EPR) 
to vanish identically despite the system being manifestly driven and dissipative.
We solve these reduced dynamics exactly for dimers and semi-analytically for general $N$-mers, 
deriving closed-form expressions for mean-squared orientational fluctuations and total EPR.
Crucially, this exact mapping demonstrates that all non-equilibrium entropy production is 
strictly partitioned into the translational (center-of-mass) sector.
Extending the framework to incorporate dipolar chemical emission reveals that while dimers retain this quasi-equilibrium structure, longer chains ($N > 2$) break it completely—generating a genuinely non-equilibrium state with non-stationary EPR. Monopolar coupling is thus both necessary and sufficient to sustain the polarized quasi-equilibrium state, establishing explicit physical boundaries for exact thermodynamic descriptions of active colloidal chains.
\end{abstract}

\maketitle

\section{Introduction}\label{sec:intro}

Active matter describes systems which are far from equilibrium: energy is injected locally and continuously at the scale of the constituent particles, and the resulting steady states generically sustain nonzero probability currents and positive entropy production~\cite{marchetti2013hydrodynamics, Cates2015, bechinger2016}. Yet a recurring and useful theme within this broader field has been the identification of special sub-classes of active systems whose steady states can nonetheless be mapped, exactly or approximately, onto equilibrium-like structures -- effective temperatures, effective free energies, or detailed-balance-obeying reduced descriptions that survive despite the microscopic irreversibility of the full dynamics \cite{vrugt2025exactly, cates2025active, ramaswamy2010mechanics, chajwa2019kepler}. Such quasi-equilibrium mappings are valuable precisely because exactly solvable non-equilibrium systems remain the exception rather than the rule: most active-matter models that admit closed-form steady-state solutions do so only in strongly reduced (mean-field, one-body, or linearized) limits, and genuine many-body non-equilibrium exact results are comparatively scarce \cite{chou2011non, foss2017solvable,lebowitz1978exact, chajwa2019kepler, bolitho2020periodic}.

Equilibrium systems owe much of their tractability to symmetry -- detailed balance is itself a symmetry statement, guaranteeing that the stationary probability current vanishes pointwise rather than merely on average. It is such symmetries that are exploited in the class of systems studied in this paper. In particular, class of experimentally realized active systems, autophoretic colloidal chains coupled through monopolar chemical (diffusiophoretic) interactions, possesses exactly this symmetry in its orientational sector, order by order in the number of monomers, once the chain geometry is held fixed \cite{Kumar2024NatComm, subramaniam2024rigid}. This is an exact statement: the reduced orientational dynamics is an exact one-dimensional gradient dynamics, and the resulting quasi-equilibrium structure -- vanishing orientational-sector EPR, an exact effective potential, a stationary distribution -- holds at any coupling strength, arbitrarily far from the linear-response regime that typically underlies equilibrium-like descriptions of active systems.

Autophoretic colloidal chains of the kind considered here have been realized experimentally as micelle-emitting droplets or Janus-type colloids linked into extended, self-propelling objects~\cite{Kumar2024NatComm, subramaniam2024rigid}, joining a broader family of extended active structures assembled via magnetic~\cite{snezhko2011magnetic}, electric~\cite{nishiguchi2018flagellar}, and other external actuation mechanisms~\cite{biswas2017linking, biswas2021rigidity}. The chemically self-interacting realization is distinguished by the presence of long-ranged, self-generated roto-translational coupling: each monomer both emits and senses a diffusing chemical field, so that the current sensed by any one monomer depends on the instantaneous configuration of the entire chain. Such chains spontaneously acquire global polar order and self-propel with a characteristic``C-shape'' topology -- a chain of essentially uniform curvature, propelling in the direction normal to its own body axis , characterized by distinct interfacial fluctuations ~\cite{Kumar2024NatComm, subramaniam2024rigid, Subramaniam2026JCP}.

The present work places this C-shape, and the broader family of dynamical steady states accessible to these chains, under the quasi-equilibrium lens described above. We show that, remarkably, a system driven far enough from equilibrium to sustain persistent self-propulsion and finite curvature nonetheless retains an exact effective-equilibrium signature in its orientational sector at every chain length we study, with all of the system's entropy production relegated to the translational (center-of-mass) degree of freedom, which we compute in closed form. We further examine the physical origin and consequence of a dipolar contribution to the chemical current -- expected on general grounds, since the micelle deposition process at the scale of an individual monomer is not obviously isotropic~\cite{Kumar2024NatComm} -- and show that this contribution, unlike the monopolar term, depends on the orientations of neighboring monomers and hence cannot be written as a gradient dynamics. We find that this loss of quasi-equilibrium structure does not manifest as a new, better-polarized steady state: a purely dipolar coupling instead produces a persistent, non-converging state, with the associated entropy production rate itself failing to reach a steady value. Monopolar coupling remains both necessary and (for the initial conditions we study) sufficient to reach the fully polarized C-shape.

\section{Model and methods}\label{sec:model}

\subsection{Equations of motion, chemical current, and excluded volume}\label{sec:eom}

We model the $i$th active colloid as a particle centered at $\mathbf{r}_i=(x_i,y_i)$, confined to move in two dimensions and self-propelling at speed $v_s$ along $\mathbf{e}_i=(\cos\theta_i,\sin\theta_i)$. The position and orientation of the $i$th particle evolve as
\begin{align}
\dot{\mathbf{r}}_i &= v_s\,\mathbf{e}_i + \mu\,\mathbf{F}_i^{b} + \mu\, \mathbf{F}_i^{ex} + \sqrt{2D_t}\,\bm\xi^t_{i}, \label{eq:pos_eom}\\
\dot{\theta}_i &= \chi_r\left(\mathbf{e}_i\times\mathbf{J}_i\right) + \sqrt{2D_r}\,\bm\xi^r_{i}, \label{eq:theta_eom}
\end{align}
where $\mu$ is the mobility, $D_t$ and $D_r$ are translational and rotational diffusion constants, and $\bm\xi^{t,r}$ are unit-variance white noises. The bonded force $\mathbf{F}_i^b=-\partial U/\partial\mathbf{r}_i$ derives from a harmonic chain potential $U=\sum_{i=1}^{N-1}k(r_{i,i+1}-r_0)^2$ of stiffness $k$ and rest length $r_0=2b$, where $b$ is the monomer radius. $\chi_r>0$ throughout unless stated otherwise, so that chemo-repulsive monomers rotate away from one another. We additionally include a soft excluded-volume repulsion $\mathbf{F}_i^{ex}$, active only between non-bonded pairs ($|i-j|\ge2$),
\begin{align}
\mathbf{F}_i^{ex} = k_{ex}\sum_{\substack{j\ne i\\|i-j|\ge2}} \frac{(r_{ex}-r_{ij})_+}{r_{ij}}\,\hat{\mathbf{r}}_{ij}, \qquad r_{ex}=2b,
\label{eq:excl_vol}
\end{align}
with $(x)_+\equiv\max(x,0)$, which prevents the chain from folding into a numerically (and physically) singular self-intersection at large curvature -- a possibility the bonded potential alone does not exclude, since it only constrains adjacent pairs.

The chemical current is obtained from the steady-state solution of the phoretic field equation $D_c\nabla^2c(\mathbf{r},t)+\sum_i c_0\,\delta(\mathbf{r}-\mathbf{r}_i)=0$, giving the monopolar current
\begin{align}
\mathbf{J}_i = \lambda_m\sum_{\substack{j\neq i}}^N \frac{\mathbf{r}_i-\mathbf{r}_j}{|\mathbf{r}_i-\mathbf{r}_j|^3}, \qquad \lambda_m\equiv\frac{c_0}{4\pi D_c},
\label{eq:mono_current}
\end{align}
where $c_0$ is the micelle emission rate and $D_c$ its diffusivity. This field equation, and hence Eq.~\eqref{eq:mono_current}, is solved in the full three-dimensional space surrounding each colloid -- Eq.~\eqref{eq:mono_current} is the resulting free-space $1/r$ Green's function gradient, $\nabla(1/r)\propto\hat{\mathbf{r}}/r^2$ -- since the diffusing chemical species is not itself confined to the plane of colloidal motion. The colloids, by contrast, are taken to move in a two-dimensional plane, as realized experimentally by gravitational or geometric confinement to a single interface or thin cell~\cite{Kumar2024NatComm}. This combination -- a genuinely three-dimensional chemical field sourced by particles whose mechanical motion is restricted to two dimensions -- is valid provided the confining geometry does not itself appreciably perturb the field at the interparticle separations of interest. 


\subsection{Initial conditions and dynamical symmetry}\label{sec:ic0}

Unless stated otherwise, all results in this work use the same initial condition: a straight chain, monomers evenly spaced by the bond rest length $2b$ along a common axis, with propulsion initially perpendicular to that axis,
\begin{align}
x_i(0) = 0, \qquad y_i(0) = (i-1)\cdot 2b, \qquad \theta_i(0) = 0, \qquad i=1,\dots,N,
\label{eq:ic0}
\end{align}
which we refer to throughout as IC0. Examples of these are shown in Fig. \ref{fig:dimer_det}(b) and \ref{fig:n5_det}(a). Equations~\eqref{eq:pos_eom}--\eqref{eq:theta_eom} are invariant under the combined operation of relabeling $i\to N+1-i$ together with $y\to-y$, $\theta\to-\theta$ (equivalently $e_x\to e_x$, $e_y\to-e_y$), since $\mathbf J_i$ depends only on relative positions and both the bonded potential $U$ and the current of Eq.~\eqref{eq:mono_current} respect this reflection.
The remarkable consequence of this choice of IC0 is that the chain geometry is mirror-symmetric about its center throughout its dynamical evolution -- $\theta_i(0)=-\theta_{N+1-i}(0)$. Thus, the exact mirror relations
\begin{align}
e_{x,i}(t) = e_{x,N+1-i}(t), \qquad e_{y,i}(t) = -e_{y,N+1-i}(t), \qquad \forall i,\ \forall t\ge0,
\label{eq:mirror_symmetry}
\end{align}
hold -- see Fig. \ref{fig:dimer_det}(a) (also \cite{subramaniam2024rigid}). These relations halve the number of independent orientational degrees of freedom, and eventually enable the equilibrium description.
This is thus a \textit{post facto} symmetry application -- we apply dynamical symmetries known from simulation and show that the system can be reduced to analytically tractable limits. In the absence of such a symmetric initial condition, an analogous mirror symmetry may still be recovered dynamically through history-dependent (trail-mediated) interactions, which have been shown to select a universal, initial-condition-independent steady state~\cite{Kumar2024NatComm, subramaniam2024rigid}; we do not pursue this history-dependent extension here, restricting attention throughout to the instantaneous, memoryless current of Eq.~\eqref{eq:mono_current}.

\subsection{Geometrical quantities}
Interior bond angles are computed directly from the simulated monomer positions via the cosine rule: writing $\mathbf{v}_k^{(1)}=\mathbf{r}_{k-1}-\mathbf{r}_k$ and $\mathbf{v}_k^{(2)}=\mathbf{r}_{k+1}-\mathbf{r}_k$ for the two bond vectors meeting at monomer $k$,
\begin{align}
\alpha_k = \arccos\!\left(\frac{\mathbf{v}_k^{(1)}\cdot\mathbf{v}_k^{(2)}}{|\mathbf{v}_k^{(1)}||\mathbf{v}_k^{(2)}|}\right), \qquad k=2,\dots,N-1.
\label{eq:bond_angle_cosine_rule}
\end{align}

\subsection{Deterministic quantities: stability and polar order}\label{sec:det_methods}

Given the deterministic fixed-point solution of any reduced equation of motion $\dot{\mathbf{X}}=\mathbf{F}(\mathbf{X})$, we linearize about it, $\delta\dot{\mathbf{X}}_i=\mathbf{M}_{ij}\delta X_j$, $\mathbf{M}_{ij}=\partial F_i/\partial X_j$, and report the eigenvalues $\lambda$ of $\mathbf{M}$; a negative real part indicates local stability, and we identify any exactly zero eigenvalue as a marginal (neutrally stable) direction. Where an analytical eigenvalue is unavailable or serves as a cross-check, we additionally obtain $\lambda_k$ numerically via central finite differences on the torque itself: $\lambda_k=[F_k(\theta_k^*+\epsilon)-F_k(\theta_k^*-\epsilon)]/2\epsilon$, evaluated at the converged fixed point with $\epsilon=10^{-6}$, $F_k$ the right-hand side of Eq.~\eqref{eq:theta_eom} for site $k$ at fixed geometry.

We also track the global polar order parameter
\begin{align}
P(t) = \frac{1}{N}\left|\sum_{i=1}^N \mathbf{e}_i(t)\right|,
\label{eq:pol_defn}
\end{align}
which interpolates between $P=0$ (isotropic/disordered orientations) and $P=1$ (perfect alignment); $P\to1$ as the chain reaches its fully polarized dynamical steady state.

\subsection{Stochastic quantities: MSD and entropy production}\label{sec:stoch_methods}

For the orientational sector we track the mean-squared displacement
\begin{align}
\langle(\Delta\theta)^2\rangle(t) = \langle(\theta(t)-\langle\theta\rangle)^2\rangle,
\label{eq:msd_defn}
\end{align}
averaged over realizations, and the associated stationary distribution $P_{ss}(\theta)$. For a generic overdamped Langevin coordinate $\dot X = F(X) + \xi$, $\langle\xi(t)\xi(t')\rangle=2D\delta(t-t')$, the steady-state entropy production rate is
\cite{peliti2021stochastic, siraishi2023, seifert_2025}:
\begin{align}
\sigma = \int dX\, \frac{J_{ss}(X)^2}{D\,P_{ss}(X)}, \qquad J_{ss}(X) = F(X)P_{ss}(X) - D\,\partial_X P_{ss}(X),
\label{eq:epr_defn}
\end{align}
which vanishes identically if and only if $J_{ss}\equiv0$ (detailed balance)~\cite{seifert2012stochastic}. Equation~\eqref{eq:epr_defn} is automatically satisfied for any one-dimensional Langevin equation with $F(X)=-\partial_X U(X)$ -- this is the precise sense in which the orientational sector is in quasi-equilibrium. The translational (center-of-mass) sector, by contrast, is driven by $\dot X_{\rm com}\propto v_s\sum_k\cos\theta_k$, a non-conservative function of the (fluctuating) orientational state, and carries the entirety of the system's dissipation.

\subsection{Simulation parameters}\label{sec:sim_params}

All stochastic simulations in this work integrate Eqs.~\eqref{eq:pos_eom}--\eqref{eq:theta_eom} using the (Ito) Euler--Maruyama scheme, with drift evaluated at the start of each timestep and noise added directly; deterministic simulations use the same scheme with the noise terms set to zero. Parameter values differ substantially between the dimensionless, fixed-geometry sections (Secs.~\ref{sec:results_det}--\ref{sec:nmer_stoch}) and the physically-scaled dipole section (Sec.~\ref{sec:dipole}); both are tabulated in full, by figure, in Appendix~\ref{app:params}.

\section{Results: deterministic dynamics}\label{sec:results_det}

\begin{figure}[h]
\centering
\includegraphics[width=0.75\textwidth]{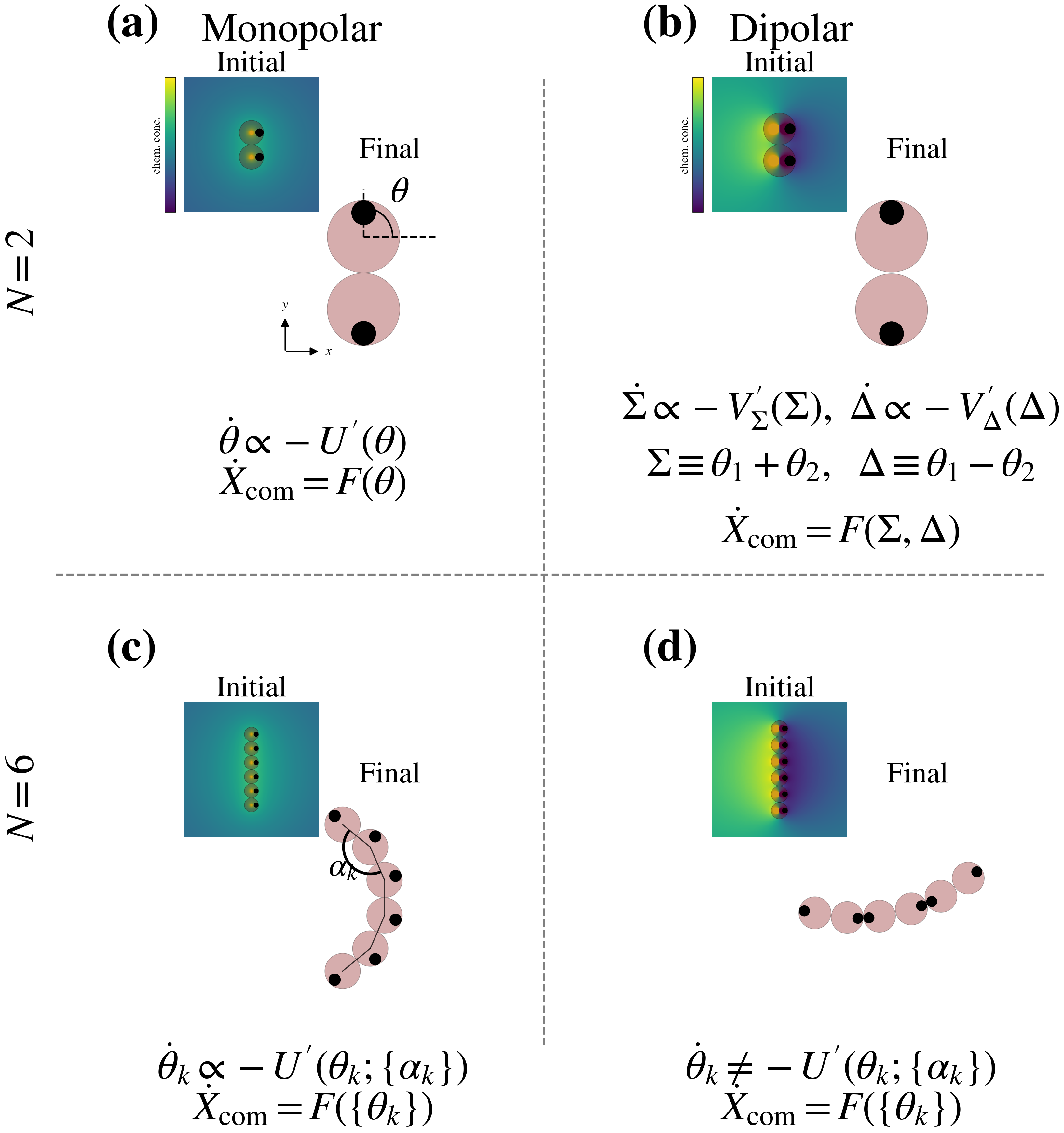}
\caption{Schematic summary of the four disntict cases studied in the paper. Columns: monopolar (a,c) vs. dipolar (b,d) coupling. Rows: dimer, $N=2$ (a,b) vs. longer chain, $N=6$ (c,d). Each panel shows the initial (chemical concentration field, colorbar shared within each panel) and final configuration, together with the reduced equation(s) of motion governing that case. (a) Monopolar dimer: the orientation $\theta$ is defined relative to the body axis as shown; $x$-$y$ coordinate convention at bottom left applies throughout. (b) Dipolar dimer: exact reduction survives only in the transformed pair $(\Sigma,\Delta)$. (c) Monopolar $N=6$: converged C-shape, with the interior bond angle $\alpha_k$ (Eq.~\eqref{eq:bond_angle_cosine_rule}) illustrated directly on the chain geometry. (d) Dipolar $N=6$: no analogous potential exists; final configuration shown is representative, not a converged fixed point. The bottom panels are distinguished by having a parametric equilibrium formulation in the $\theta$ sector (in the monopolar case), whereas in the upper panel a full equilibrium description of this sector is available. We note that in all panels, the $X_{com}$ dynamics adds an additional non-equilibrium structure to the dynamics. We note that the display of the chemical concentration fields in the insets are for illustrative purposes only; all simulations in this work use particle-based simulations of \ref{eq:pos_eom}.}
\label{fig:summary_schematic}
\end{figure}

It is first useful to summarize the overall structure of the results to follow. Figure~\ref{fig:summary_schematic} presents all four combinations of coupling (monopolar vs. dipolar) and chain length ($N=2$ vs. a longer $N=6$ chain) that organize the rest of this section and Sec.~\ref{sec:dipole}. For monopolar coupling, both the dimer (a) and the general $N$-mer (c) reduce, at fixed geometry, to an orientational dynamics that is an exact gradient dynamics of a single-particle-like potential $U$, such that $\delta\dot\theta\propto-U'(\delta\theta)$ for the dimer, $\dot\theta_k\propto-U'(\theta_k;\{\alpha_k\})$ for each mode of the longer chain, parametrized by the instantaneous bond angles $\{\alpha_k\}$ (the same $\alpha_k$ defined in Eq.~\eqref{eq:bond_angle_cosine_rule}, illustrated here directly on the converged $N=6$ C-shape) -- with all translational motion, $\dot X_{\rm cm}=F(\theta)$ or $F(\{\theta_k\})$, sourced by but not feeding back into this orientational sector. Dipolar coupling preserves this exact gradient structure only for the dimer (b), where it survives in the transformed pair $(\Sigma,\Delta)\equiv(\theta_1+\theta_2,\theta_1-\theta_2)$ rather than in $\theta_1$ directly; for the longer chain (d) no such reduction exists in general, $\dot\theta_k\ne-U'(\theta_k;\{\alpha_k\})$, and the orientational dynamics is no longer a gradient dynamics at all. This progression -- from an exact single-variable potential, to an exact two-variable one, to no potential whatsoever -- is evident in Secs.~\ref{sec:results_det}--\ref{sec:dipole}. Video 1 in the \cite{SupplementalMaterial} summarizes the results of the deterministic dynamics for these cases. We now go through each cases in detail.

\subsection{Dimers}\label{sec:dimers_det}

\begin{figure}[h]
\centering
\includegraphics[width=0.85\textwidth]{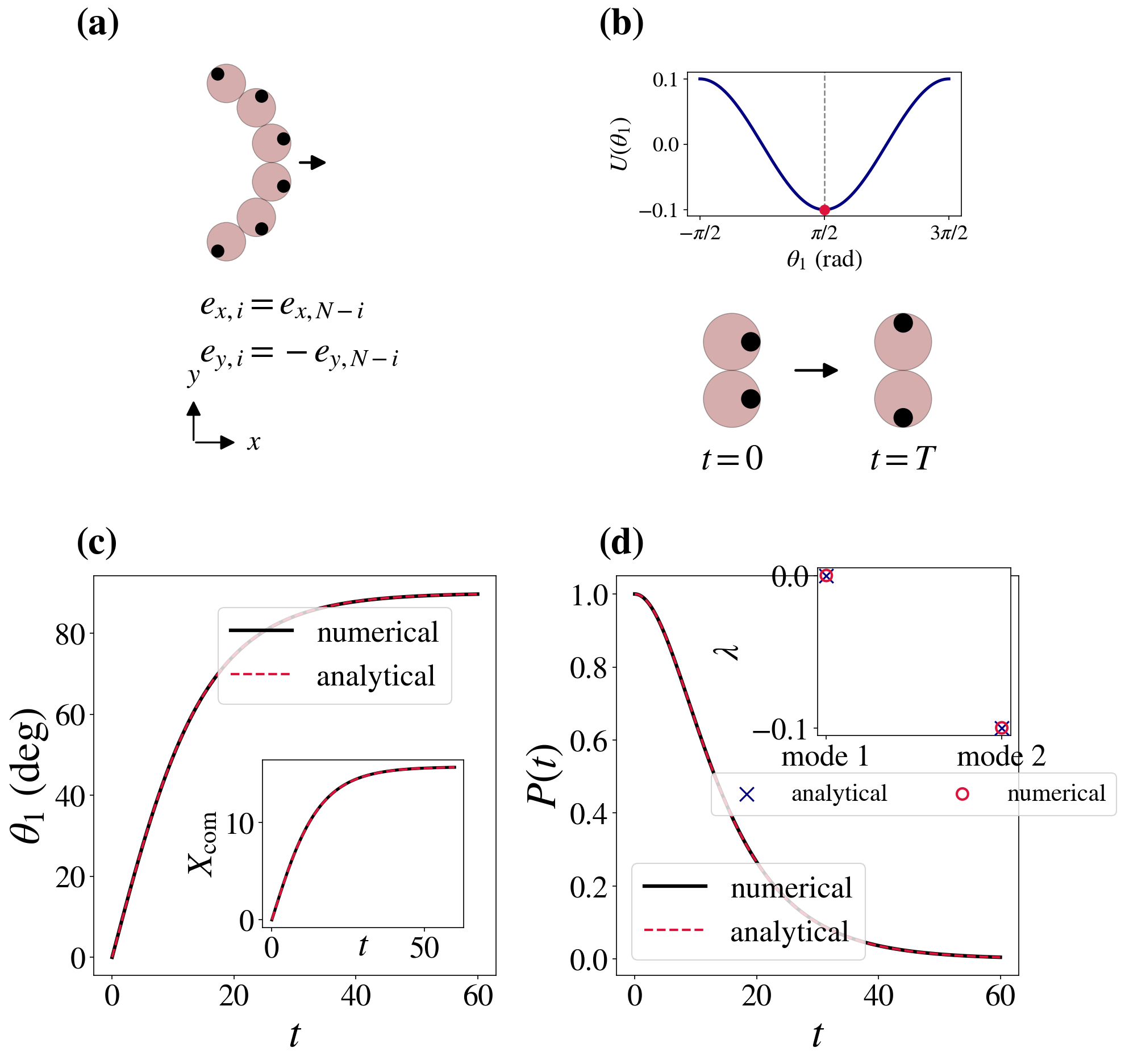}
\caption{Deterministic quasi-equilibrium solution for the dimer ($N=2$). (a) Representative $N=6$ C-shape configuration, illustrating the mirror symmetry $\theta_k=-\theta_{N+1-k}$ about the body axis and the bond-angle labeling convention $\alpha_k$; propulsion direction indicated by the arrow, with coordinate axes shown at bottom left. Simulations of any Nmers under (\ref{eq:pos_eom}) with initial conditions (\ref{eq:ic0}) will always exhibit such a symmetry. (b) Initial ($t=0$) and final ($t=T$) dimer configurations, connected by an arrow indicating the transition (not a trajectory). Inset: the effective potential $U(\theta_1)$ of Eq.~\eqref{eq:dimer_eom}, with its minimum at $\theta_1^*=\pi/2$ marked. (c) Orientation $\theta_1(t)$: numerical integration of the full coupled dynamics (Eqs.~\eqref{eq:pos_eom}--\eqref{eq:theta_eom}) vs. the closed-form solution, Eq.~\eqref{eq:dimer_soln}. Inset: center-of-mass displacement $X_{\rm com}(t)$, numerical vs. Eq.~\eqref{eq:dimer_xcom}. (d) Global polar order $P(t)$, numerical vs. analytical, decaying from $P(0)=1$ (the straight initial condition) to $P\to0$ as the dimer halts (Sec.~\ref{sec:dimers_det}). Inset: the two LSA eigenvalues, analytical ($-\chi_r R$ and the exactly marginal $0$) vs. numerical (finite-difference Jacobian, Sec.~\ref{sec:det_methods}).}
\label{fig:dimer_det}
\end{figure}

For $N=2$, the empirically and analytically confirmed symmetry $\theta_2(t)=-\theta_1(t)$ holds throughout the dynamics, reducing the system to the single independent angle $\theta_1$ \cite{subramaniam2024rigid}. See also appendix \ref{sec:dimerAPP} for stability analysis of the dynamical systems of dimers.
At fixed bond length $r_0=2b$, equations (\ref{eq:pos_eom}) and (\ref{eq:theta_eom}) can be simplified via the symmetries of (\ref{eq:mirror_symmetry}). The reduced equation of motion for the orientational sector is
\begin{align}
\dot\theta_1 = -\frac{a}{2}\cos\theta_1, \qquad a \equiv \frac{\chi_r\lambda_m}{(4b)^2},
\label{eq:dimer_eom}
\end{align}
which is derivable from the effective potential $U(\theta_1)=\tfrac{a}{2}\sin\theta_1$ via $\dot\theta_1=-\partial_{\theta_1}U$ (Fig.~\ref{fig:dimer_det}(b), inset): an explicit, closed-form confirmation of the quasi-equilibrium structure claimed above. Equation~\eqref{eq:dimer_eom} has the exact solution, from $\theta_1(0)=\theta_0$,
\begin{align}
\theta_1(t) = \phi + 2\arctan\!\left[\tan\!\left(\tfrac{\theta_0-\phi}{2}\right)e^{-\chi_r R\,t}\right], \qquad R = \frac{\lambda_m}{(2b)^2},
\label{eq:dimer_soln}
\end{align}
which for $\theta_0=0$ reduces to $\theta_1(t)=2\arctan(e^{-(a/2)t})-\pi/2$, the Gudermannian function of $-(a/2)t$ -- the same function that describes the motion of an undamped nonlinear pendulum along its separatrix~\cite{goldstein2002classical}, with $\theta_1^*=-\pi/2$ playing the role of the (marginally reached, infinite-time) inverted equilibrium; Fig.~\ref{fig:dimer_det}(c) confirms this solution against direct numerical integration of the full coupled dynamics. Using the kinematic identity $\dot X_{\rm com}=v_s\cos\theta_1$ (which follows directly from $\theta_2=-\theta_1$), Eq.~\eqref{eq:dimer_soln} integrates in closed form to
\begin{align}
X_{\rm com}(t) = X_{\rm com}(0) + \frac{v_s}{\chi_r R}\,\arctan\!\big[\sinh(\chi_r R\,t)\big],
\label{eq:dimer_xcom}
\end{align}
using $\cos\theta_1(t)={\rm sech}(\chi_r R\,t)$ for the $\theta_0=0$ solution -- the dimer's center of mass approaches a finite total displacement rather than propelling indefinitely (Fig.~\ref{fig:dimer_det}(c), inset). Thus, the monopolar dimer always comes to a halt~\cite{Kumar2024NatComm, subramaniam2024rigid}, reflected directly in the global polar order $P(t)$ decaying from $P(0)=1$ (the initially straight, aligned chain) to $P\to0$ as the two monomers settle into antiparallel orientations (Fig.~\ref{fig:dimer_det}(d)).

Linearizing Eq.~\eqref{eq:dimer_eom} about $\theta_1^*=-\pi/2$ gives eigenvalue $\lambda_\theta=-\chi_r R$; the marginal $X_{\rm com}$ direction (Eq.~\eqref{eq:dimer_xcom} depends only on $t$, not on a restoring force) contributes a second, exactly zero eigenvalue -- both confirmed numerically via the finite-difference Jacobian of Sec.~\ref{sec:det_methods} (Fig.~\ref{fig:dimer_det}(d), inset). The dimer is thus marginally stable - orientationally trapped and translationally marginally free.

\subsection{Generic $N$-mers}\label{sec:nmer_det}

For a symmetric $N$-mer at fixed bond angles $\{\alpha_j\}$, the reflection symmetry $\theta_k=-\theta_{N+1-k}$ again reduces the independent orientational degrees of freedom to $\lfloor N/2\rfloor$. Since $\mathbf{J}_k$ depends only on the (fixed) geometry, each independent $\theta_k$ obeys its own decoupled, one-dimensional equation
\begin{align}
\dot\theta_k = \chi_r R_k(\{\alpha_j\})\sin\big[\phi_k(\{\alpha_j\}) - \theta_k\big],
\label{eq:nmer_eom}
\end{align}
with amplitude $R_k$ and phase $\phi_k$ fixed, at any given time, by the instantaneous geometry -- confirming that the quasi-equilibrium structure of Sec.~\ref{sec:dimers_det} generalizes to arbitrary $N$, site by site, provided the geometry itself is held fixed (the \emph{parametric} quasi-equilibrium referred to throughout this work). Each $\theta_k$ has the same closed-form solution as Eq.~\eqref{eq:dimer_soln}, with $(\chi_r R, \phi)\to(\chi_r R_k,\phi_k)$, and a stable fixed point at $\theta_k^*=\phi_k$.

For the trimer, $R(\alpha)$ and $\phi(\alpha)$ follow directly from the geometry. Fix monomer 2 at the origin with monomer 1 at unit distance along a reference direction, and monomer 3 at the same bond length, turned by the interior angle $\alpha$ (bond length set to $2b\equiv1$ here; restored below). Writing $s\equiv\sin(\alpha/2)$, $c\equiv\cos(\alpha/2)$, the chord length to the non-bonded neighbor (monomer 3, as seen from monomer 1) is $2s$ -- the base of the isoceles triangle with equal sides $1$ and apex angle $\alpha$ -- and the corresponding unit vector is $(-s,-c)$. Summing the bonded contribution (unit vector $(-1,0)$, unit distance) and this non-bonded contribution into $\mathbf{J}_1=K\sum_{j\ne1}\hat{\mathbf r}_{1j}/r_{1j}^2$ gives
\begin{align}
\mathbf{J}_1/K = \left(-1,\,0\right) + \frac{(-s,-c)}{4s^2} \equiv -(A,\,B), \qquad
A = 1+\frac{1}{4s}, \qquad B = \frac{c}{4s^2}.
\label{eq:trimer_AB}
\end{align}
Squaring and using $c^2=1-s^2$, the $1/(16s^2)$ terms cancel exactly, leaving the closed form
\begin{align}
R(\alpha) = \sqrt{A^2+B^2} = \sqrt{1+\frac{1}{2\sin(\alpha/2)}+\frac{1}{16\sin^4(\alpha/2)}}, \qquad \phi(\alpha) = \pi+\arctan(B/A),
\label{eq:trimer_Rphi}
\end{align}
verified to match direct numerical evaluation of $\mathbf{J}_1$ on the explicit geometry across the full physical range $\alpha\in(0,\pi)$. Restoring the bond length $2b$ rescales $R\to\lambda_m R/(2b)^2$. Since the stable orientational fixed point is $\theta_1^*=\phi(\alpha)$ (Sec.~\ref{sec:nmer_det}), the corresponding LSA eigenvalue is, in closed form,
\begin{align}
\lambda_\theta(\alpha) = -\chi_r R(\alpha) = -\chi_r\sqrt{A^2+B^2}.
\label{eq:trimer_eigenvalue}
\end{align}

This closed form is not special to $N=3$. Repeating the same construction for $N=4$ (edge monomer, uniform interior bond angle $\alpha$ at both junctions) introduces a third-neighbor chord, $r_{14}=|1-4s^2|$, which itself factors cleanly ($8s^4-6s^2+1=(1-2s^2)(1-4s^2)$), giving for $\alpha>60^\circ$ (our working range; $\alpha=60^\circ$ is a genuine geometric degeneracy at which the third-neighbor chord vanishes and monomers 1 and 4 coincide) the closed-form contribution $(\cos\alpha,-\sin\alpha)/(4s^2-1)^2$, verified against direct evaluation including near this singular point. A fully general, compact closed form for arbitrary $N$ is obtainable in exactly this way. For $N=5$ (Sec.~\ref{sec:nmer_det}) we therefore evaluate $R_k(\alpha),\phi_k(\alpha)$ via the same construction; the closed-form expressions for these are written in Appendix~\ref{app:n5_closed_form}.

\begin{figure}[h]
\centering
\includegraphics[width=0.95\textwidth]{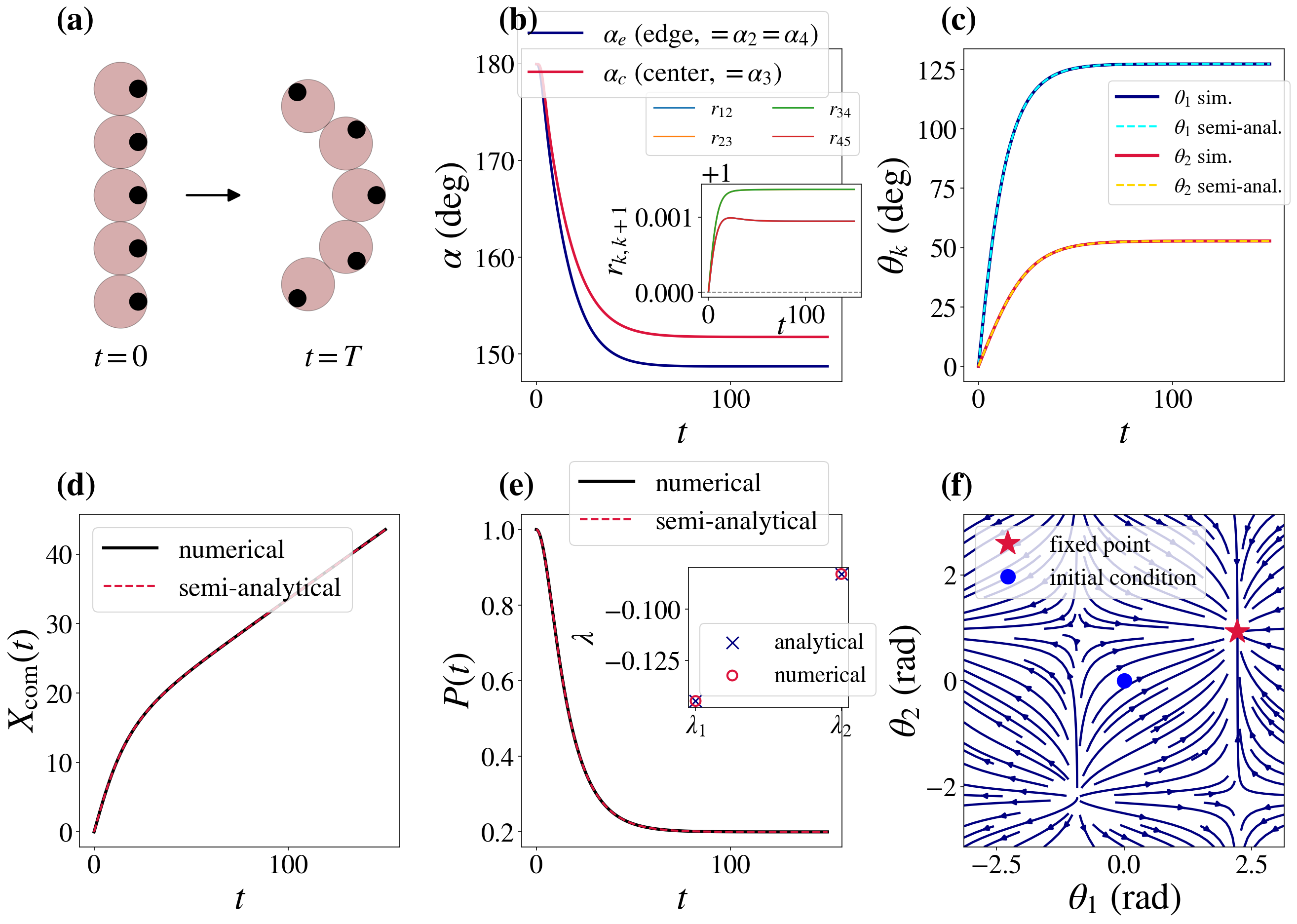}
\caption{Deterministic quasi-equilibrium solution for the pentamer ($N=5$). (a) Initial ($t=0$) and final ($t=T$) pentamer configurations. (b) Bond angles $\alpha_e$ (edge, $=\alpha_2=\alpha_4$) and $\alpha_c$ (center, $=\alpha_3$) vs. time, computed via the cosine rule (Eq.~\eqref{eq:bond_angle_cosine_rule}) on the simulated positions; the two converge to distinct values, confirming the uniform-bond-angle approximation used in earlier sections is not exact for the true self-consistent geometry. Inset: bond lengths $r_{k,k+1}$, converging to $r_0=2b$. (c) Orientations $\theta_1,\theta_2$: full simulation vs. the semi-analytical solution obtained by re-integrating the reduced EOM (Eq.~\eqref{eq:nmer_eom}) using the same simulated positions -- the two are exact to numerical precision, direct evidence for the quasi-equilibrium theorem of Sec.~\ref{sec:nmer_det}. (d) Center-of-mass displacement $X_{\rm com}(t)$, numerical vs. semi-analytical (obtained by integrating $\dot X_{\rm com}=v_s\langle\cos\theta_k\rangle_k$ against the semi-analytical $\theta_k(t)$). (e) Global polar order $P(t)$, numerical vs. semi-analytical. Inset: the two nonzero LSA eigenvalues $\lambda_1,\lambda_2$, analytical (Eq.~\eqref{eq:n5_eigenvalues}) vs. numerical (finite-difference Jacobian). (f) Phase portrait in $(\theta_1,\theta_2)$ (radians), streamlines of the reduced dynamics with the converged fixed point marked in red (initial condition in blue).}
\label{fig:n5_det}
\end{figure}

For the pentamer, the reflection symmetry leaves two independent bond angles, $\alpha_e\equiv\alpha_2=\alpha_4$ (edge junctions) and $\alpha_c\equiv\alpha_3$ (center junction), and two independent orientational modes, $\theta_1$ (edge) and $\theta_2$. $R_1(\alpha_e,\alpha_c)$ and $R_2(\alpha_e,\alpha_c)$ follow in closed form from Eqs.~\eqref{eq:n5_A1B1_general}--\eqref{eq:n5_A2B2_general} of Appendix~\ref{app:n5_closed_form} via $R_k=\sqrt{A_k^2+B_k^2}$, giving the closed-form LSA eigenvalues
\begin{align}
\lambda_1(\alpha_e,\alpha_c) = -\chi_r R_1(\alpha_e,\alpha_c), \qquad \lambda_2(\alpha_e,\alpha_c) = -\chi_r R_2(\alpha_e,\alpha_c),
\label{eq:n5_eigenvalues}
\end{align}
and, via Eq.~\eqref{eq:dimer_soln} with $(\chi_r R,\phi)\to(\chi_r R_k,\phi_k)$, the closed-form solution $\theta_k(t)$ for each mode -- now valid for independently chosen edge and center bond angles, not restricted to a uniform geometry. At the representative angle used in Fig.~\ref{fig:n5_det} ($\alpha_e=\alpha_c=150^\circ$), this gives $R_1=1.452$, $\phi_1=-172.10^\circ$ and $R_2=0.815$, $\phi_2=-98.65^\circ$, so $\lambda_1=-1.452\,\chi_r$ and $\lambda_2=-0.815\,\chi_r$: as with the dimer, the $X_{\rm com}$ direction contributes one additional, exactly zero eigenvalue. Figure~\ref{fig:n5_det} confirms this full closed-form solution -- bond angles (b), orientations (c), center-of-mass displacement (d), polar order (e), and the two LSA eigenvalues (e, inset) -- against direct numerical integration of the full coupled dynamics; panel (f) shows the corresponding phase portrait in $(\theta_1,\theta_2)$.

We conclude this section by noting that the same quasi-equilibrium description established for the dimer persists for general $N$ - each independent orientational mode is an exact one-dimensional gradient dynamics, parametrized by (and decoupled given) the instantaneous bond-angle geometry, while the translational sector, driven nonconservatively by the collective orientational state, is solely responsible for breaking this structure.

\subsection{Polar order and stability for generic N-mers}\label{sec:scaling_det}

\begin{figure}[h]
\centering
\includegraphics[width=0.7\textwidth]{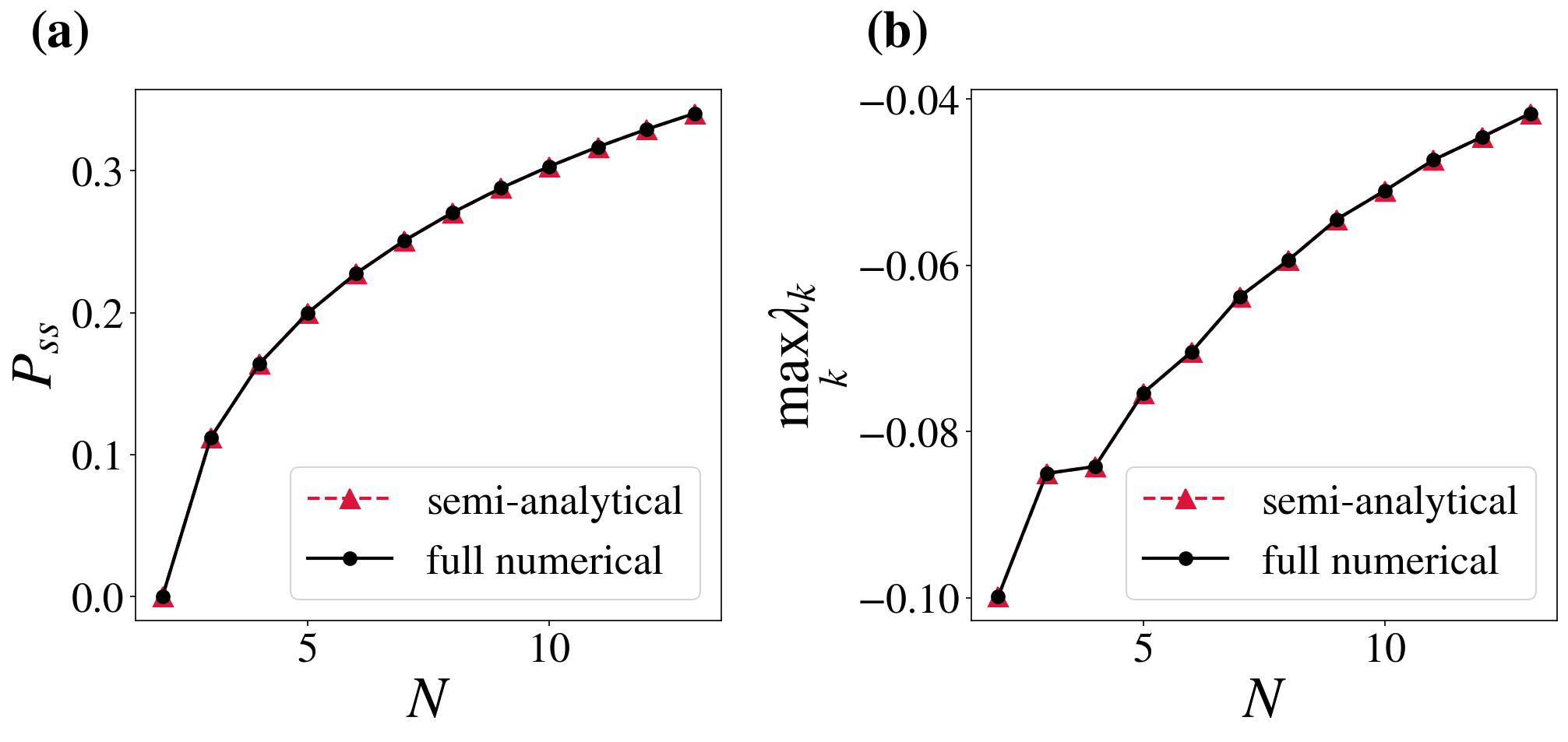}
\caption{Growth of polar order and maximum eigenvalue with chain length $N$, from a genuine self-consistent $N$-sweep (each $N$ run to its own converged geometry via the full coupled dynamics). (a) Steady-state global polar order $P_{ss}$ vs. $N$, growing sub-linearly from $P_{ss}=0$ at $N=2$ (the dimer halts exactly, Sec.~\ref{sec:dimers_det}). (b) Least-stable (weakest, closest to marginal) LSA eigenvalue, $\max_k\lambda_k$, vs. $N$, increasing monotonically toward zero -- longer chains are more polarized (a) but less stable to orientational perturbations (b). In both panels, ``semi-analytical" (evaluating $R_k$ from Eq.~\eqref{eq:mono_current} and the fixed-point condition $\theta_k=\phi_k$ directly on the converged simulated geometry) and ``full numerical" (from the simulated trajectory and a finite-difference Jacobian) agree.}
\label{fig:scaling_det}
\end{figure}

Figure~\ref{fig:scaling_det} confirms this: direct numerical evaluation, via a genuine self-consistent $N$-sweep from $N=2$ to $N=13$ (each $N$ run to its own converged geometry through the full coupled dynamics, not the uniform-bond-angle approximation of Sec.~\ref{sec:nmer_det}), shows $P_{ss}(N)$ growing sub-linearly from $P_{ss}=0$ at $N=2$ (the dimer halts exactly) to $P_{ss}\approx0.34$ at $N=13$, saturating rather than continuing to grow linearly with chain length, consistent with the corresponding scaling reported in our earlier work~\cite{Kumar2024NatComm,subramaniam2024rigid} (Fig.~\ref{fig:scaling_det}(a)). The least-stable (weakest, closest-to-marginal) LSA eigenvalue $\max_k\lambda_k$ increases essentially monotonically toward zero over the same range, from $-0.0998$ at $N=2$ to $-0.0417$ at $N=13$, with only a slight non-monotonicity between $N=3$ ($-0.0850$) and $N=4$ ($-0.0842$) (Fig.~\ref{fig:scaling_det}(b)); the qualitative picture is that longer chains are more polarized and stiffer, but less stable to orientational perturbations. In both panels, the semi-analytical prediction (evaluating $R_k$ and the fixed-point condition $\theta_k=\phi_k$ directly on the converged simulated geometry) and the full numerical result agree to within numerical precision at every $N$ tested, confirming the quasi-equilibrium theorem holds independent of chain length.

A plausible mechanism for the sub-linear growth follows from decomposing $P_{ss}$ itself. Writing $NP_{ss}^2=\big|\sum_i\mathbf{e}_i\big|^2=\sum_{i,j}\cos(\theta_i-\theta_j)$ and separating the diagonal ($i=j$) from the off-diagonal terms, $P_{ss}^2 = \frac{1}{N} + \frac{1}{N^2}\sum_{i\ne j}\cos(\theta_i-\theta_j)$,
the first term alone would give $P_{ss}\sim1/\sqrt N\to0$; it is entirely the second, correlation term that must grow with $N$ to produce the observed sub-linear \emph{increase} rather than decay. For a chain of locally constant curvature (bond angle varying slowly along the chain, $\theta_k\approx k\,\Delta\theta$ for some effective per-bond angular increment $\Delta\theta$), the sum above is exactly the same geometric-series structure used throughout Sec.~\ref{sec:nmer_det} and Appendix~\ref{app:n5_closed_form}, and evaluates in closed form to the standard array-factor result $P_{ss} = \frac{\big|\sin(N\Delta\theta/2)\big|}{N\big|\sin(\Delta\theta/2)\big|}$,
verified to reproduce $NP_{ss}^2$ exactly for a linear angular progression. If $\Delta\theta$ were independent of $N$, the expression would eventually \emph{decrease} with $N$ once the total angular span $N\Delta\theta$ exceeds $\pi$ (destructive interference, the chain curling back on itself); the observed monotonic increase therefore requires $\Delta\theta$ itself to shrink with $N$. This is directly consistent with the eigenvalue data of Fig.~\ref{fig:scaling_det}(b): since $R_k$ sets the local curvature-driving torque and decreases with $N$ (each monomer's current is progressively diluted by the growing number of distant, weakly-$1/r^2$-coupled neighbors), the self-consistent steady-state bond angle $\alpha^*(N)$ approaches $\pi$ as $N$ grows, i.e. $\Delta\theta(N)\to0$. If the total angular span $N\Delta\theta(N)$ saturates to a finite limit rather than growing linearly, the expression correspondingly saturates toward a nonzero asymptotic value rather than decaying -- consistent with, and offering a mechanistic account of, the vanishing-curvature limit $\kappa\to0$ as $N\to\infty$ already reported for this system~\cite{subramaniam2024rigid}. A full analysis of this would require the self-consistent scaling of $\alpha^*(N)$ itself, obtained here only via the externally-fixed-$\alpha$ reduction (Sec.~\ref{sec:nmer_det}), not solved for; a first-principles derivation of $\Delta\theta(N)$ is left to future work.

\section{Results: stochastic dynamics}\label{sec:results_stoch}

\subsection{Dimers}\label{sec:dimers_stoch}

Linearizing Eq.~\eqref{eq:dimer_eom} about $\theta_1^*=-\pi/2$ with additive rotational noise gives the orientational mean-squared displacement, measured from the (deterministically) equilibrated starting point,
\begin{align}
\langle(\Delta\theta_1)^2\rangle(t) = \frac{2D_r}{a}\left(1-e^{-at}\right), \qquad a=\chi_r R,
\label{eq:dimer_msd}
\end{align}
an exact Ornstein-Uhlenbeck-type trapping of the orientational coordinate about $\theta_1^*$, saturating at plateau $2D_r/a$. The exact (not merely linearized) stationary distribution, obtained directly from $P_{ss}(\theta_1)\propto\exp[-U(\theta_1)/D_r]$ using the effective potential of Sec.~\ref{sec:dimers_det}, is the von Mises distribution
\begin{align}
P_{ss}(\theta_1) = \frac{\exp[k\cos(\theta_1-\theta_1^*)]}{2\pi I_0(k)}, \qquad k = \frac{a}{2D_r},
\label{eq:dimer_vonmises}
\end{align}
with $I_n$ the modified Bessel function of the first kind. Since $\theta_1$'s own dynamics is an exact gradient dynamics, the orientational-sector EPR (Eq.~\eqref{eq:epr_defn} applied to $\theta_1$ itself) vanishes identically, at any $D_r$ -- the quasi-equilibrium signature stated in Sec.~\ref{sec:intro}. The full system EPR, carried entirely by $X_{\rm com}$ via the kinematic relation of Sec.~\ref{sec:dimers_det}, evaluates in closed form to
\begin{align}
\sigma_{\rm dimer} = \frac{2v_s^2}{D_t}\,\langle\cos^2\theta_1\rangle_{ss} = \frac{2v_s^2}{D_t}\cdot\frac{I_1(k)}{k\,I_0(k)},
\label{eq:dimer_epr}
\end{align}
which is manifestly non-negative and finite for any $D_r>0$, and approaches this same small, non-vanishing value in the long-time dynamic limit as well: even the deterministic-path (leading-order) prediction for $\sigma(t)$, corrected for the fluctuation contribution about the relaxing trajectory, settles to the Bessel steady-state value of Eq.~\eqref{eq:dimer_epr} rather than to zero -- fluctuations about the halted configuration $\theta_1^*=-\pi/2$ dissipate at any finite $D_r$, even though the mean drift itself vanishes there. Only the orientational-sector EPR $\sigma_\theta$, not the full $\sigma$, goes to exactly zero, both at steady state and dynamically as $\theta_1(t)\to\theta_1^*$ (see Video 2 of SM \cite{SupplementalMaterial}).\\

\subsection{$N$-mers (trimer)}\label{sec:nmer_stoch}

\begin{figure}[t]
\includegraphics[width=0.85\textwidth]{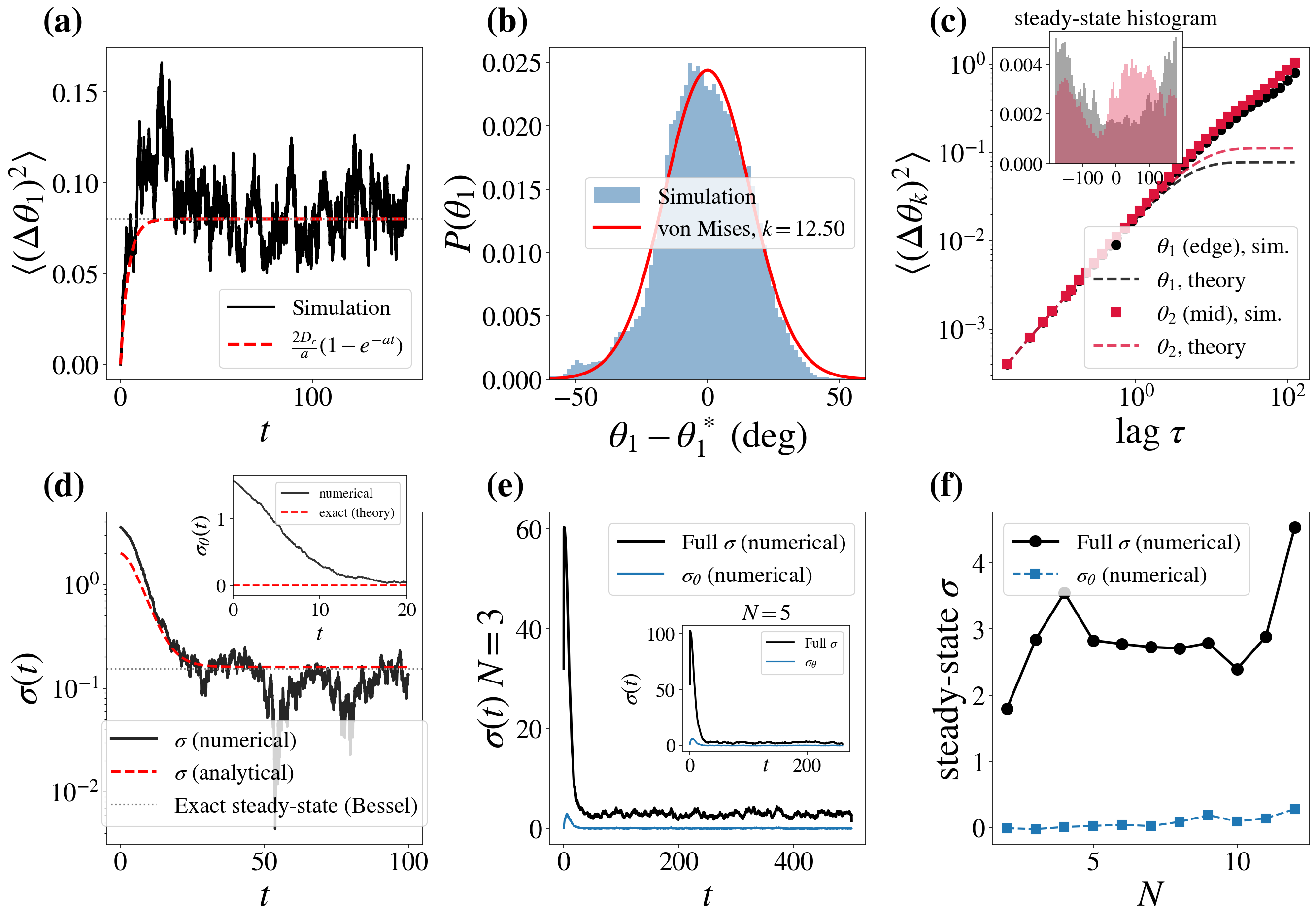}
\caption{(a) Dimer MSD: numerical vs. analytical (Eq.~\eqref{eq:dimer_msd}), showing agreement at all accessible times. (b) Dimer steady-state distribution: histogram of $\theta_1-\theta_1^*$ vs. the exact von Mises form (Eq.~\eqref{eq:dimer_vonmises}). (c) Trimer MSD: numerical vs. semi-analytical, showing agreement only at short lag and clear, growing deviation at long lag (log-log axes); inset shows the steady-state histogram of $\theta_k-\theta_k^*$ (numerical only -- the fixed-geometry theoretical prediction is omitted, as it does not match well once bond-angle fluctuation is present, consistent with the long-lag MSD deviation in the main panel). (d) Dimer EPR $\sigma(t)$: numerical and analytical (leading order with fluctuation correction), both approaching the small, non-vanishing Bessel steady-state value of Eq.~\eqref{eq:dimer_epr} rather than zero; inset shows the orientational-sector EPR $\sigma_\theta(t)$, which does vanish exactly, both curves. (e) Trimer EPR: full $\sigma(t)$ and orientational-sector $\sigma_\theta(t)$, both numerical (no analytical prediction available once bond-angle dynamics is included, Sec.~\ref{sec:nmer_stoch}); inset shows the same two quantities for $N=5$, confirming both are genuinely nonzero in steady state. (f) Steady-state EPR (full $\sigma$ and orientational $\sigma_\theta$) vs. chain length, $N=2$--$12$ (numerical).}
\label{fig:msd_hist}
\label{fig:epr}
\end{figure}

The same linearization applied to Eq.~\eqref{eq:nmer_eom}, at fixed geometry, predicts an OUP-trapped MSD of the same functional form as Eq.~\eqref{eq:dimer_msd} for each independent $\theta_k$, with $a\to a_k=\chi_r R_k(\alpha)$, and a von Mises steady state $P_{ss}(\theta_k)$ of the same form as Eq.~\eqref{eq:dimer_vonmises}. Direct comparison against genuine simulation (bond angles evolving under their own, unconstrained dynamics rather than held externally fixed) confirms this prediction only at short lag times; at long times the true MSD continues to grow, with no sign of saturation, exceeding the semi-analytical plateau by an order of magnitude or more by the end of the accessible simulation window. The steady-state distribution of $\theta_k$ itself is correspondingly broader than the fixed-geometry von Mises prediction and is not well described by it (Fig.~\ref{fig:msd_hist}(c), inset). The origin of this discrepancy is genuine bond-angle fluctuation: once the geometry is allowed to evolve rather than being held externally fixed, each $\theta_k$ is no longer a closed, autonomous one-dimensional system. We note that the underlying equations of motion, Eqs.~\eqref{eq:pos_eom}--\eqref{eq:theta_eom}, are invariant under a global rotation (rotating every position and every $\theta_k$ by the same constant angle leaves the dynamics unchanged, since only relative positions and relative orientations enter $\mathbf{F}_i^b$ and $\mathbf{J}_i$); this symmetry offers a plausible qualitative account of the observed long-time growth -- free diffusion of the chain's overall orientation, superposed on a genuinely trapped local (relative-to-body-axis) angle -- though we have not verified this explicitly (e.g. via a corresponding zero eigenvalue of the full, non-reduced linearization) and present it here as a plausible interpretation rather than a demonstrated result. A full treatment of this coupled bond-angle/orientation problem is beyond the scope of the present work.

The EPR computed from the fixed-geometry reduction, by the same Eq.~\eqref{eq:dimer_epr} generalized to include the cross terms $\langle\cos\theta_k\rangle\langle\cos\theta_j\rangle$ arising from the independent (but not identically distributed) $\theta_k$, does \emph{not} vanish at $N=3$ even in the small-$D_r$ limit, in contrast to the dimer: the coherent (mean-drift) contribution to the translational EPR, $(\textstyle\sum_k\cos\phi_k)^2$, is exactly zero for $N=2$ by the halting symmetry of Sec.~\ref{sec:dimers_det}, but is generically nonzero for $N\ge3$, where no analogous exact cancellation is guaranteed. Direct simulation of the dynamic EPR (both full $\sigma$ and orientational-sector $\sigma_\theta$) confirms this: unlike the dimer, both settle to a genuinely nonzero steady-state value for the trimer and for $N=5$ alike (Fig.~\ref{fig:msd_hist}(e)).

It is also of interest to note the behaviour of the steady-state EPR with chain length: starting from a value that vanishes as $D_r\to0$ at $N=2$, $\sigma_{ss}$ grows for larger $N$ as this coherent contribution becomes more significant -- longer chains, and hence more strongly polarized, faster-propelling objects (Sec.~\ref{sec:scaling_det}), dissipate correspondingly more, though the direct simulation shows a non-monotonic, noisy trend rather than a clean growth law across $N=2$--$12$ (Fig.~\ref{fig:msd_hist}(f)). We note that the fixed-geometry prediction of Eq.~\eqref{eq:general_epr} shows good quantitative agreement with direct simulation for the trimer, but does \emph{not} agree well with numerics for $N=5$: this is a direct consequence of the bond-angle fluctuations themselves not being accounted for in the fixed-geometry reduction (Sec.~\ref{sec:nmer_stoch}), which grow more significant as $N$ increases and are not captured by treating $R_k,\phi_k$ as constants evaluated at a single geometry. These plots for $N=6$ is show in Video 3 of the SM \cite{SupplementalMaterial}.\\

\section{Dipolar chemical field}\label{sec:dipole}

The monopolar current of Eq.~\eqref{eq:mono_current} follows from the assumption of an isotropic, point-like emission of chemical species at each monomer. This is an idealization: in the emulsion-droplet realization of these chains~\cite{Kumar2024NatComm}, the process by which filled micelles are deposited at an individual monomer is not obviously symmetric, and a leading correction to Eq.~\eqref{eq:mono_current} takes the form of a dipolar emission term,
\begin{align}
\mathbf{J}_i^{\rm total} = \mathbf{J}_i^{\rm mono} - \lambda_d\sum_{\substack{j\ne i}} \frac{(\hat{\mathbf{e}}_j\cdot\hat{\mathbf{r}}_{ij})\hat{\mathbf{r}}_{ij} - \hat{\mathbf{e}}_j/3}{r_{ij}^3},
\label{eq:dipolar_current}
\end{align}
where $\hat{\mathbf{r}}_{ij}=(\mathbf{r}_i-\mathbf{r}_j)/r_{ij}$. Unlike the monopolar term, Eq.~\eqref{eq:dipolar_current} depends explicitly on the orientation $\hat{\mathbf{e}}_j$ of every \emph{other} monomer $j$, and hence cannot in general be written as $-\partial_{\theta_i}U(\{\theta_k\})$ for any single-particle potential $U$: the argument of Secs.~\ref{sec:results_det}--\ref{sec:results_stoch}, that $\mathbf{J}_i$ depends only on position and therefore the orientational dynamics is an exact gradient dynamics, no longer applies once $\lambda_d\ne0$. This is the sense in which dipolar emission breaks the quasi-equilibrium structure identified above.
To quantify the resulting shape we use the polarization--axis order parameter
\begin{align}
\Psi \equiv \mathbf{P}\cdot\hat{\mathbf{t}}, \qquad \mathbf{P}\equiv\frac{1}{N}\sum_k\hat{\mathbf{e}}_k, \qquad \hat{\mathbf{t}}\equiv\frac{\mathbf{r}_N-\mathbf{r}_1}{|\mathbf{r}_N-\mathbf{r}_1|},
\label{eq:psi_defn}
\end{align}
the global polarization vector $\mathbf{P}$ (the same quantity as in Ref.~\cite{subramaniam2024rigid}, Eq.~(8)) projected onto the unit vector along the chain's body axis (its end-to-end chord). For a chain in the C-shape configuration -- monomer propulsion everywhere perpendicular to the local bond, i.e., radially outward along a circular arc of bend angle $\Phi$ -- direct integration gives $\mathbf{P}=\Phi^{-1}(\sin\Phi,\,1-\cos\Phi)$ and a chord $\mathbf{r}_N-\mathbf{r}_1 \propto (\cos\Phi-1,\,\sin\Phi)$, whose dot product vanishes identically for \emph{any} bend angle $\Phi$: $\Psi=0$ \emph{exactly} is the C-shape signature, independent of how strongly curled the arc is. A nonzero $|\Psi|$ instead signals that $\mathbf{P}$ has a component along the chain's own axis -- e.g. a straight, uniformly-aligned flock, for which $\mathbf{P}$ is parallel to $\hat{\mathbf{t}}$ and $|\Psi|=|\mathbf{P}|$. Both quantities are compared directly across the full $(\lambda_m,\lambda_d)$ plane in App.~\ref{app:dipole_phase_diagram}.

\subsection{Equilibrium dynamics of the dipolar dimer}\label{sec:dipole_dimer_eq}

\begin{figure}[t]
\centering
\includegraphics[width=0.95\textwidth]{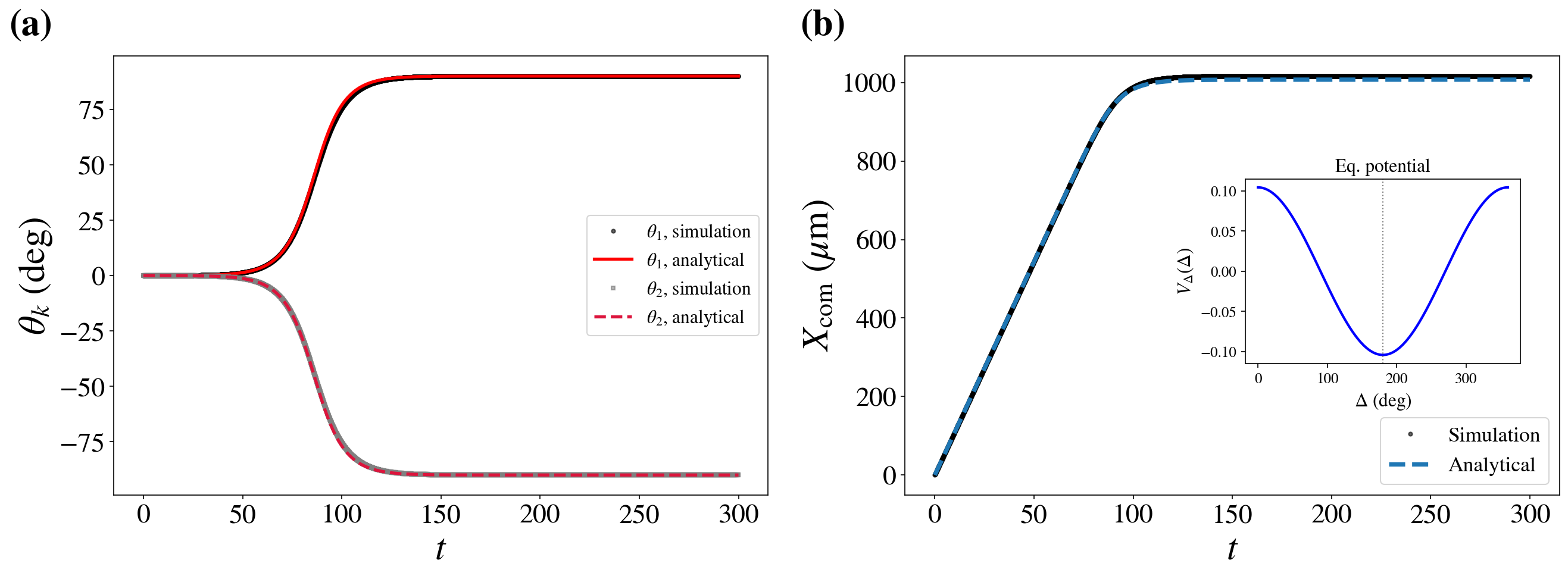}
\caption{Dipolar dimer, IC0, $\lambda_m=0$. (a) $\theta_1,\theta_2$ vs. $t$: direct simulation vs. the analytical solution of Eqs.~\eqref{eq:dipdimer_sigma}--\eqref{eq:dipdimer_delta}. (b) $X_{\rm com}$ vs. $t$: simulation vs. analytical; inset shows the effective potential $V_\Delta(\Delta)=a_{\rm dip}\cos\Delta$ governing the (slower) $\Delta$ mode, whose minimum at $\Delta=\pi$ sets the halted configuration.}
\label{fig:dipole_dimer_eq}
\end{figure}

The general argument of Sec.~\ref{sec:dipole} -- that $\mathbf{J}_i$ depending explicitly on $\hat{\mathbf{e}}_j$ for $j\ne i$ breaks the single-particle gradient-flow structure of Secs.~\ref{sec:results_det}--\ref{sec:results_stoch} -- is correct in general, but admits a clean exception at $N=2$. Writing $\Sigma\equiv\theta_1+\theta_2$ and $\Delta\equiv\theta_1-\theta_2$, the dipolar-only ($\lambda_m=0$) dimer's equations of motion decouple \emph{exactly} (not merely to linear order, and for arbitrary $\theta_1,\theta_2$, not only near the fixed point):
\begin{align}
\dot\Sigma &= -b_\Sigma\sin\Sigma, & b_\Sigma&=\lambda_d/8b^3, \label{eq:dipdimer_sigma}\\
\dot\Delta &= -a_{\rm dip}\sin(\Delta-\pi), & a_{\rm dip}&=\lambda_d/(24b^3)=b_\Sigma/3, \label{eq:dipdimer_delta}
\end{align}
 Each is an independent, exact one-dimensional gradient dynamics -- $\Sigma$ relaxing to $\Sigma^*=0$, $\Delta$ to $\Delta^*=\pi$, at rates differing by a factor of 3 -- so the dipolar dimer is, in this precise sense, \emph{also} a quasi-equilibrium system: both $\Sigma$ and $\Delta$ individually satisfy detailed balance, exactly as $\theta_1$ does for the monopolar dimer (Sec.~\ref{sec:dimers_stoch}). We verified Eqs.~\eqref{eq:dipdimer_sigma}--\eqref{eq:dipdimer_delta} directly against the full multi-particle engine, both on and off the eventual fixed point (Fig.~\ref{fig:dipole_dimer_eq}(a)). Since $r_{ij}$ is set by the (essentially rigid) bond, the center-of-mass velocity, $\dot X_{\rm com}=v_s\cos(\Sigma/2)\cos(\Delta/2)$ (Fig.~\ref{fig:dipole_dimer_eq}(b)), vanishes exactly at the fixed point ($\Delta^*=\pi\Rightarrow\cos(\Delta^*/2)=0$): the dipolar dimer halts, by the same mechanism as the monopolar one -- fluctuation about a configuration of exactly zero mean drift, not a residual drift itself.

The key structural difference from the monopolar dimer is that the quasi-equilibrium structure here lives in the \emph{pair} $(\Sigma,\Delta)$ rather than in $\theta_1$ alone: it is a genuinely two-body reduction (there is no single-particle potential $U(\theta_1)$ analogous to Sec.~\ref{sec:dimers_det}), which is also why it does not survive to $N\ge3$, where three or more mutually-coupled orientations admit no analogous pair of decoupled combinations in general. Because both $\Sigma$ and $\Delta$ are nonetheless exact, independent 1D gradient dynamicss, the stochastic consequences carry over with only minor modification: the orientational-sector EPR vanishes identically (both sectors individually satisfy detailed balance), the stationary distributions of $\Sigma$ and $\Delta$ are again von Mises, and the full EPR is carried entirely by $X_{\rm com}$, vanishing in the steady-state, and as $D_r\to0$ exactly as in Sec.~\ref{sec:dimers_stoch}. We do not reproduce the stochastic figures here, as they are qualitatively identical in form to Fig.~\ref{fig:msd_hist}(a),(b),(d) for the monopolar dimer; the exact expressions, which differ from the monopolar case in detail (a two-mode rather than single-mode relaxation, and a different Bessel-function combination for the EPR), are derived in App.~\ref{app:dipole_dimer_eq}.

\subsection{Deterministic solution for Nmers}
\begin{figure}[t]
\centering
\includegraphics[width=0.7\textwidth]{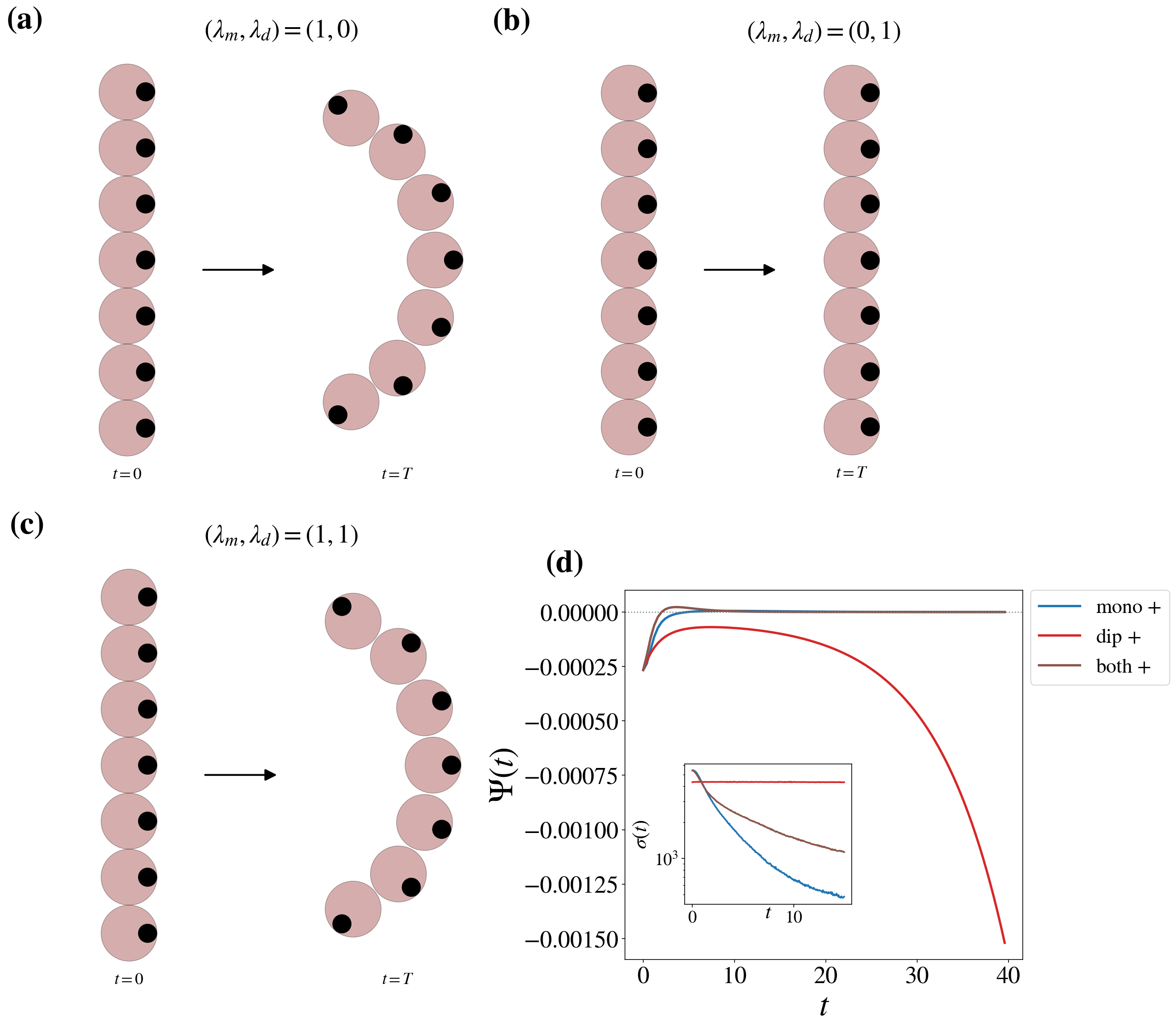}
\caption{Dipole contribution, IC0, deterministic. (a)--(c) Snapshots at $t=0$ and $t=T$ for, respectively, purely monopolar ($\lambda_m,\lambda_d)=(1,0)$, purely dipolar $(0,1)$, and both couplings present $(1,1)$ (all in units of the respective base values; arrow indicates the transition, not a trajectory). (d) $\Psi(t)$, Eq.~\eqref{eq:psi_defn}, for the three cases; inset shows the corresponding stochastic entropy production rate $\sigma(t)$ (Sec.~\ref{sec:dipole} following).}
\label{fig:dipole_ic0}
\end{figure}

For IC0, a purely monopolar coupling ($\lambda_m>0$, $\lambda_d=0$) reaches the fully polarized C-shape, $\Psi\to0$ exactly, confirming both the deterministic solution of Sec.~\ref{sec:results_det} and the exact perpendicularity condition of Eq.~\eqref{eq:psi_defn}, directly against genuine (non-fixed-geometry) simulation (Fig.~\ref{fig:dipole_ic0}(a),(d)). A purely dipolar coupling ($\lambda_m=0$, $\lambda_d\ne0$) does not reach any static configuration: $\Psi(t)$ fails to converge even at long times, instead exhibiting sustained, non-monotonic drift with no fixed limiting value (Fig.~\ref{fig:dipole_ic0}(b),(d)). When both couplings are present with $\lambda_m>0$, the monopolar contribution is sufficient to recover the fully polarized state, $\Psi\to0$ exactly, essentially unaffected by the presence of $\lambda_d$ (Fig.~\ref{fig:dipole_ic0}(c),(d)); the dipolar contribution is therefore neither necessary for, nor does it enhance, C-shape formation at this initial condition -- it is monopolar coupling alone that both produces and stabilizes the polarized state.

\subsection{Stochastic solution}
Adding small rotational and translational noise to the same dynamics and computing the resulting entropy production rate via the multi-degree-of-freedom generalization of Eq.~\eqref{eq:epr_defn} (summed over all positional and orientational coordinates, with the position-sector force Jacobian obtained numerically), we find that purely monopolar coupling reaches a genuine steady-state EPR, settling to a stable plateau at long times (Fig.~\ref{fig:dipole_ic0}(d), inset). Purely dipolar coupling does not: the EPR itself fails to reach a steady value, instead fluctuating persistently -- the entropy-production signature of the same non-convergent behavior identified in the deterministic solution above. For more, see videos 4 and 5 of the SM \cite{SupplementalMaterial}.\\
\section{Discussion}\label{sec:discussion}

We have identified an exact quasi-equilibrium structure in the orientational dynamics of monopolar, chemically self-interacting active colloidal chains: at any fixed bond-angle geometry, every independent orientational degree of freedom obeys a one-dimensional gradient dynamics, detailed balance holds exactly, and the associated entropy production vanishes --  as an exact consequence of the symmetries of the dynamics. This holds for the dimer exactly and for general $N$-mers semi-analytically, with all system dissipation carried by the translational sector, which we compute in closed form for the dimer and semi-analytically for larger $N$.

This quasi-equilibrium structure of the generic $N$-mer is explicitly \emph{parametric}: it is a semi-analytical deterministic solution in the sense that the bond angles need to be kept track of for the full solution. Allowing the bond angles to fluctuate -- as they must in any genuine, unconstrained simulation -- couples the orientational and positional sectors and both quantitatively (mean-squared displacement growing without bound at long times) and qualitatively (a nonzero orientational-sector EPR once bond-angle coupling is present) breaks the exact structure established under the fixed-geometry reduction. We interpret the residual long-time orientational growth as free rotational diffusion of the whole-chain orientation -- an emergent zero mode of the rotationally invariant model -- superposed on a genuinely bounded, locally trapped angle relative to the chain's own (fluctuating) body axis; a systematic theory of this coupled problem remains open.

A second, and in our view more surprising, mode of breaking this structure is furnished by dipolar chemical emission, motivated by the physically expected asymmetry of micelle deposition at the individual-monomer scale. Because the dipolar current depends on neighboring monomers' orientations, not merely their positions, it cannot in general be written in gradient form, and the orientational quasi-equilibrium theorem of Secs.~\ref{sec:results_det}--\ref{sec:results_stoch} does not extend to it. It is notable how this structure is broken: rather than producing an alternative static (if non-equilibrium) steady-state, a purely dipolar coupling instead produces sustained, non-monotonic drift with no fixed limiting configuration, with an entropy production rate that itself never settles. This is consistent with a general expectation for driven, non-gradient dynamical systems, which (unlike gradient dynamicss, which must always terminate at a fixed point) are free to sustain genuinely time-dependent, non-converging behavior. Monopolar coupling, precisely because it retains the gradient structure, remains both necessary and (for the symmetric initial condition we study) sufficient to reach the fully polarized C-shape; dipolar coupling, by contrast, opens the door to a qualitatively different class of non-equilibrium steady state without itself producing a new static one.

Potential future work includes the study of stochastic thermodynamics in this system -- either the optimal work required to drive the chain between configurations in finite time \cite{SchmiedlSeifert2007}, or the corresponding optimal heat dissipated, for which active systems admit a genuine optimal (finite) protocol duration achieving the best trade-off between external and internal dissipation \cite{DavisProesmansFodor2024}. In addition, the study of chemo-attractive ($B$-type, with those studied here being the ``$A$-type'') and mixed ($BAB$/$ABA$) interactions are also possible due to the existence of symmetries mentioned here \cite{subramaniam2024rigid}. The universality of the C-shape (independence from our chioce of initial conditions) -- established to be due to the presence of trails \cite{Kumar2024NatComm, subramaniam2024rigid} for the monopolar case -- can also be extended to the dipolar case.

\appendix

\section{Simulation parameters and calibration}\label{app:params}

This appendix collects every parameter value used to produce the results and figures of Secs.~\ref{sec:results_det}--\ref{sec:dipole}, listed per figure in Table~\ref{tab:params}. Parameter values differ substantially between the dimensionless, fixed-geometry sections (Secs.~\ref{sec:results_det}--\ref{sec:nmer_stoch}) and the physically-scaled dipole section (Sec.~\ref{sec:dipole}), which use two independent calibrations rather than a shared unit system; parameters common to all rows within each group are stated once in the table caption rather than repeated per row.

\begin{table}[h]
\centering
\scriptsize
\setlength{\tabcolsep}{3pt}
\begin{tabular}{@{}lccccccl@{}}
\hline\hline
\textbf{Figure} & $N$ & Coupling & $T$ & $dt$ & $N_{\rm real}$ & IC & \textbf{Remarks} \\
\hline
Fig.~\ref{fig:dimer_det} & 2 & $K=1/8$ & 60 & $5\times10^{-5}$ & -- & straight & Det.; $a_{\rm dimer}=0.25$ \\
Fig.~\ref{fig:n5_det} & 5 & $K=0.1$ & 150 & $5\times10^{-5}$ & -- & straight & Det. \\
Fig.~\ref{fig:scaling_det} & 2--13 & $K=0.1$ & 250 & $5\times10^{-5}$ & -- & straight & Det., self-consistent $N$-sweep \\
Fig.~\ref{fig:msd_hist}(a,b) & 2 & $K=1/8$ & 60--350 & $5\times10^{-4}$ & 30 & straight & Stochastic \\
Fig.~\ref{fig:msd_hist}(c--e) & 3, 5 & $K=0.1$ & 150--500 & $5\times10^{-4}$ & 30 & straight & Stochastic \\
Fig.~\ref{fig:msd_hist}(f) & 2--12 & $K=0.1$ & 260 & $5\times10^{-4}$ & 15 & straight & Indicative only \\
Fig.~\ref{fig:dipole_dimer_eq} & 2 & $\lambda_d=\lambda_m b$ & 300 & 0.01 & -- & IC0 & Dipolar only \\
Fig.~\ref{fig:dipole_ic0} & 7 & $\lambda_m,\lambda_d\in\{0,{\rm base}\}$ & 300 & 0.01 & 15 & IC0 & -- \\
Fig.~\ref{fig:phase_diagram_ic0} & 7 & $20\times20$ grid$^*$ & 200 & 0.01 & -- & IC0 & -- \\
\hline\hline
\end{tabular}
\caption{Simulation parameters by figure. Global, fixed-across-all-panels quantities not shown per row: the dimensionless sections (Figs.~\ref{fig:dimer_det}--\ref{fig:msd_hist}) use $\chi_r=v_s=\mu=1$, $b=0.5$, $l_{sp}=1$, $k_{sp}=1200$, $D_r=0.01$, $D_t=0.05$; the dipole sections (Figs.~\ref{fig:dipole_dimer_eq}--\ref{fig:phase_diagram_ic0}) use real units with $v_s=b/2.3$, $\mu=1/(6\pi b)$, $b=25\times10^{-6}$, $l_{sp}=2b$, $k_{sp}=225b$, $\lambda_m^{\rm base}=10^6b^3\times0.1$, $D_r=10^{-4}\lambda_m/b^2$, $D_t=10^{-4}v_sb$.}
\label{tab:params}
\end{table}

\section{Closed form expression of $R_k$ for $N=5$}\label{app:n5_closed_form}

We work with two independent bond angles, $\alpha_e\equiv\alpha_2=\alpha_4$ (the edge junctions, equal to one another by the reflection symmetry $\theta_k=-\theta_{N+1-k}$) and $\alpha_c\equiv\alpha_3$ (the center junction, self-paired under the same symmetry and hence unconstrained). As for the $N=4$ case above, the chain geometry follows from explicit turning angles $\delta_e=\pi-\alpha_e$, $\delta_c=\pi-\alpha_c$ at the respective junctions, giving four distinct non-bonded chord lengths, according to which junction(s) are spanned:
\begin{align}
r_A^2 &= 2-2\cos\alpha_e && \text{(one edge junction)}, \notag\\
r_D^2 &= 2-2\cos\alpha_c && \text{(one center junction)}, \notag\\
r_B^2 &= 3-2\cos\alpha_c-2\cos\alpha_e+2\cos(\alpha_c+\alpha_e) && \text{(edge+center junctions, either order)}, \notag\\
r_C^2 &= 4-2\cos\alpha_c-4\cos\alpha_e+4\cos(\alpha_c+\alpha_e)-2\cos(\alpha_c+2\alpha_e) && \text{(edge+center+edge)}.
\label{eq:n5_general_chords}
\end{align}
That $r_B$ is the same expression regardless of whether the edge or center junction is crossed first is a direct consequence of the chord-length formula depending only on the sequence of junction \emph{types} spanned, not on absolute chain position -- the same structural fact underlying the chord-length construction of Sec.~\ref{sec:nmer_det} in the uniform-angle case above. Evaluating $\mathbf{J}_k/\lambda_m\equiv A_k+iB_k$ term by term, the three independent sites of the $N=5$ chain are:
\begin{align}
(A_1,B_1) &= \frac{(-1,0)}{1} + \frac{(\cos\alpha_e-1,\,-\sin\alpha_e)}{r_A^3} + \frac{(\cos\alpha_e-\cos(\alpha_c+\alpha_e)-1,\;-\sin\alpha_e+\sin(\alpha_c+\alpha_e))}{r_B^3} \notag\\
&\quad+ \frac{(\cos\alpha_e-\cos(\alpha_c+\alpha_e)+\cos(\alpha_c+2\alpha_e)-1,\;-\sin\alpha_e+\sin(\alpha_c+\alpha_e)-\sin(\alpha_c+2\alpha_e))}{r_C^3},
\label{eq:n5_A1B1_general}
\end{align}
\begin{align}
(A_2,B_2) &= \frac{(1,0)}{1} + \frac{(\cos\alpha_e,\,-\sin\alpha_e)}{1} + \frac{(\cos\alpha_e-\cos(\alpha_c+\alpha_e),\;-\sin\alpha_e+\sin(\alpha_c+\alpha_e))}{r_D^3} \notag\\
&\quad+ \frac{(\cos\alpha_e-\cos(\alpha_c+\alpha_e)+\cos(\alpha_c+2\alpha_e),\;-\sin\alpha_e+\sin(\alpha_c+\alpha_e)-\sin(\alpha_c+2\alpha_e))}{r_B^3},
\label{eq:n5_A2B2_general}
\end{align}
\begin{align}
(A_3,B_3) &= \frac{(1-\cos\alpha_e,\,\sin\alpha_e)}{r_A^3} + \frac{(-\cos\alpha_e,\,\sin\alpha_e)}{1} + \frac{(-\cos(\alpha_c+\alpha_e),\,\sin(\alpha_c+\alpha_e))}{1} \notag\\
&\quad+ \frac{(-\cos(\alpha_c+\alpha_e)+\cos(\alpha_c+2\alpha_e),\;\sin(\alpha_c+\alpha_e)-\sin(\alpha_c+2\alpha_e))}{r_A^3}.
\label{eq:n5_A3B3_general}
\end{align}
Each has been verified against direct numerical evaluation of $\mathbf{J}_k$ on the explicit geometry, for both equal and genuinely unequal $(\alpha_e,\alpha_c)$ pairs. As a consistency check, setting $\alpha_e=\alpha_c\equiv\alpha$ throughout reduces $r_A\to2\sin(\alpha/2)$, $r_B\to|1-4\sin^2(\alpha/2)|$ (the trimer result), confirmed both symbolically and numerically. $R_k=\sqrt{A_k^2+B_k^2}$ and $\phi_k=\arg(A_k+iB_k)$ follow as before, now as genuine functions of both independent bond angles.

\section{Derivation of the entropy production rate}\label{app:epr_derivation}

\subsection{Orientational sector: exact vanishing}

For a single fixed-geometry mode $\theta_k$ obeying $\dot\theta_k=F_k(\theta_k)+\sqrt{2D_r}\xi_k$ with $F_k(\theta_k)=\chi_r R_k\sin(\phi_k-\theta_k)=-\partial_{\theta_k}U_k(\theta_k)$, $U_k(\theta_k)=-\chi_r R_k\cos(\phi_k-\theta_k)$ (Sec.~\ref{sec:nmer_det}), the stationary Fokker--Planck current is $J_{ss}(\theta_k)=F_k(\theta_k)P_{ss}(\theta_k)-D_r\partial_{\theta_k}P_{ss}(\theta_k)$. Steady state requires $\partial_{\theta_k}J_{ss}=0$, so $J_{ss}$ is a constant independent of $\theta_k$; since $\theta_k$ is an angle, $P_{ss}$ and $U_k$ are both $2\pi$-periodic, and integrating $J_{ss}=F_kP_{ss}-D_r\partial_{\theta_k}P_{ss}$ once around the full period forces this constant to vanish identically (a nonzero constant current would make $P_{ss}(\theta_k)=P_{ss}(0)e^{-U_k(\theta_k)/D_r}-\text{(nonperiodic term)}\propto J_{ss}$ fail to close periodically). With $J_{ss}\equiv0$, Eq.~\eqref{eq:epr_defn} gives $\sigma_{\theta_k}=0$ exactly, at any $D_r>0$ -- the orientational-sector quasi-equilibrium signature stated in Sec.~\ref{sec:intro}, and the direct consequence of $F_k$ being an exact gradient on a periodic domain.

The stationary distribution itself, from $J_{ss}\equiv0$, is the von Mises form $P_{ss}(\theta_k)\propto\exp[k_k\cos(\theta_k-\phi_k)]$, $k_k\equiv\chi_r R_k/D_r$, used throughout Sec.~\ref{sec:dimers_stoch}--\ref{sec:nmer_stoch}.

\subsection{Translational sector: general $N$}

Internal bonded forces cancel exactly in $X_{\rm com}=\tfrac{1}{N}\sum_i x_i$ (Newton's-third-law pairwise cancellation along the chain), leaving $\dot X_{\rm com}=\tfrac{v_s}{N}\sum_k\cos\theta_k+\sqrt{2D_t}\,\tfrac{1}{N}\sum_i\xi_{t,i}$. The $N$ independent unit-strength noises combine, by standard variance addition, into a single effective noise of strength $D_{\rm com}=D_t/N$: writing $\tfrac{1}{N}\sum_i\xi_{t,i}=\sqrt{1/N}\,\xi_{\rm com}$ with $\xi_{\rm com}$ itself unit-strength, $\dot X_{\rm com}=\tfrac{v_s}{N}\sum_k\cos\theta_k+\sqrt{2D_t/N}\,\xi_{\rm com}$. Since $F_{X_{\rm com}}$ does not depend on $X_{\rm com}$ itself (only on the orientational state), the $\langle\partial_{X_{\rm com}}F\rangle$ term of the general (multi-degree-of-freedom) entropy-production formula vanishes identically, leaving
\begin{align}
\sigma_{\rm trans} = \frac{\langle F_{X_{\rm com}}^2\rangle_{ss}}{D_{\rm com}} = \frac{v_s^2}{N D_t}\Big\langle\Big(\sum_k\cos\theta_k\Big)^2\Big\rangle_{ss} = \frac{v_s^2}{ND_t}\left[\sum_k\langle\cos^2\theta_k\rangle_{ss} + \sum_{k\ne j}\langle\cos\theta_k\rangle_{ss}\langle\cos\theta_j\rangle_{ss}\right],
\label{eq:general_epr}
\end{align}
using that different modes $\theta_k,\theta_j$ are statistically independent under the fixed-geometry reduction (each driven by its own independent noise $\xi_k$), so $\langle\cos\theta_k\cos\theta_j\rangle_{ss}=\langle\cos\theta_k\rangle_{ss}\langle\cos\theta_j\rangle_{ss}$ for $k\ne j$. The two required von Mises moments, for $P_{ss}(\theta_k)\propto\exp[k_k\cos(\theta_k-\phi_k)]$, are obtained by writing $\cos\theta_k=\cos[(\theta_k-\phi_k)+\phi_k]$, expanding via the angle-addition formula, and using the standard von Mises moments $\langle\cos[n(\theta_k-\phi_k)]\rangle_{ss}=I_n(k_k)/I_0(k_k)$ together with the Bessel recurrence $I_0(k_k)-I_2(k_k)=(2/k_k)I_1(k_k)$:
\begin{align}
\langle\cos\theta_k\rangle_{ss} = \cos\phi_k\,\frac{I_1(k_k)}{I_0(k_k)}, \qquad
\langle\cos^2\theta_k\rangle_{ss} = \cos^2\phi_k - \cos(2\phi_k)\,\frac{I_1(k_k)}{k_k\,I_0(k_k)},
\label{eq:vonmises_moments}
\end{align}
both verified by direct numerical quadrature against the exact von Mises distribution across the full range of $\phi_k$. Equations~\eqref{eq:general_epr}--\eqref{eq:vonmises_moments}, together with $k_k=\chi_r R_k/D_r$ and $R_k,\phi_k$ from Sec.~\ref{sec:nmer_det} (Appendix~\ref{app:n5_closed_form} for $N=5$), give the full-system EPR $\sigma=\sigma_{\rm trans}$ (since $\sigma_{\theta_k}=0$ for every mode) at fixed geometry, for any $N$.

\paragraph{Dimer limit.} For $N=2$, $\phi_1=-\pi/2$, $\phi_2=-\phi_1=\pi/2$ (the $\theta_2=-\theta_1$ symmetry), so $\cos\phi_1=\cos\phi_2=0$ and $\cos(2\phi_1)=\cos(2\phi_2)=\cos(\mp\pi)=-1$: the cross term $\langle\cos\theta_1\rangle\langle\cos\theta_2\rangle$ vanishes exactly (each factor is individually zero -- the halting symmetry of Sec.~\ref{sec:dimers_det}), and both diagonal moments reduce to $\langle\cos^2\theta_k\rangle_{ss}=I_1(k)/(kI_0(k))$ with the common $k=\chi_r R/D_r$. Equation~\eqref{eq:general_epr} then collapses to $\sigma_{\rm dimer}=\tfrac{v_s^2}{2D_t}\cdot2\cdot\tfrac{I_1(k)}{kI_0(k)}=\tfrac{2v_s^2}{D_t}\tfrac{I_1(k)}{kI_0(k)}$, recovering Eq.~\eqref{eq:dimer_epr} exactly. For general $N\ge3$, $\phi_k\ne\pm\pi/2$ in general, the cross terms are generically nonzero, and Eq.~\eqref{eq:general_epr} does not reduce to this simple form -- consistent with the qualitative statement of Sec.~\ref{sec:nmer_stoch} that the coherent translational contribution vanishes only for $N=2$.

\subsection{Numerical method of EPR computation}\label{app:epr_numerical}

Where entropy production is estimated directly from simulated stochastic trajectories, we integrate Eqs.~\eqref{eq:pos_eom}--\eqref{eq:theta_eom} with the standard (Ito) Euler--Maruyama scheme -- drift evaluated at the start of each timestep, noise (additive) added directly. Entropy production rate (EPR) is a thermodynamic (Sekimoto/stochastic-energetics) quantity defined via a \emph{Stratonovich} product of force and velocity, $\delta Q\propto F \cdot \dot X$, whereas a direct average of $F^2$ along an Ito-simulated trajectory yields the \emph{Ito}-product average instead \cite{Sekimoto1998}. The two differ by a calculable correction: for $\dot X=F(X)+\sqrt{2D}\,\xi$ (Ito), the standard Stratonovich--Ito conversion for the functional $F(X_t)\cdot \dot X_t$ gives
\begin{align}
\langle F\cdot \dot X\rangle = \langle F\,\dot X\rangle_{\rm Ito} + D\langle\partial_XF\rangle = \langle F^2\rangle + D\langle\partial_XF\rangle,
\label{eq:strat_ito_conversion}
\end{align}
using that $\langle F\dot X\rangle_{\rm Ito}=\langle F^2\rangle$ exactly (the noise term is non-anticipating and has zero mean at the resolution of a single Ito step). Dividing by $D$, the entropy production rate for a single coordinate is therefore estimated, at each saved timestep and averaged over independent realizations, as
\begin{align}
\sigma_X = \frac{\langle F^2\rangle}{D} + \langle\partial_XF\rangle,
\label{eq:epr_numerical}
\end{align}
summed over all positional and orientational degrees of freedom for the full-system $\sigma(t)$ (or restricted to the orientational coordinates alone for $\sigma_\theta(t)$). For the orientational sector, $\partial_\theta F_\theta=\partial_\theta(\mathbf e\times\mathbf J)=-\mathbf e\cdot\mathbf J$ is obtained analytically, using that $\mathbf J$ itself does not depend on $\theta$; for the translational sector, $\partial_xF_x,\partial_yF_y$ are obtained by central finite difference on the bonded-plus-excluded-volume force, as in Sec.~\ref{sec:sim_params}. This numerical estimator is used to produce Fig.~\ref{fig:msd_hist}(d) (dimer, full $\sigma(t)$ and orientational $\sigma_\theta(t)$), Fig.~\ref{fig:msd_hist}(e) (trimer, main panel; $N=5$, inset), Fig.~\ref{fig:msd_hist}(f) (steady-state EPR vs. $N$), and the insets of Fig.~\ref{fig:dipole_ic0}(d).

\section{Stability analysis of dimer dynamics}
 \label{sec:dimerAPP}

\begin{figure}[t] 
    \centering
    \includegraphics[width=.64\textwidth]{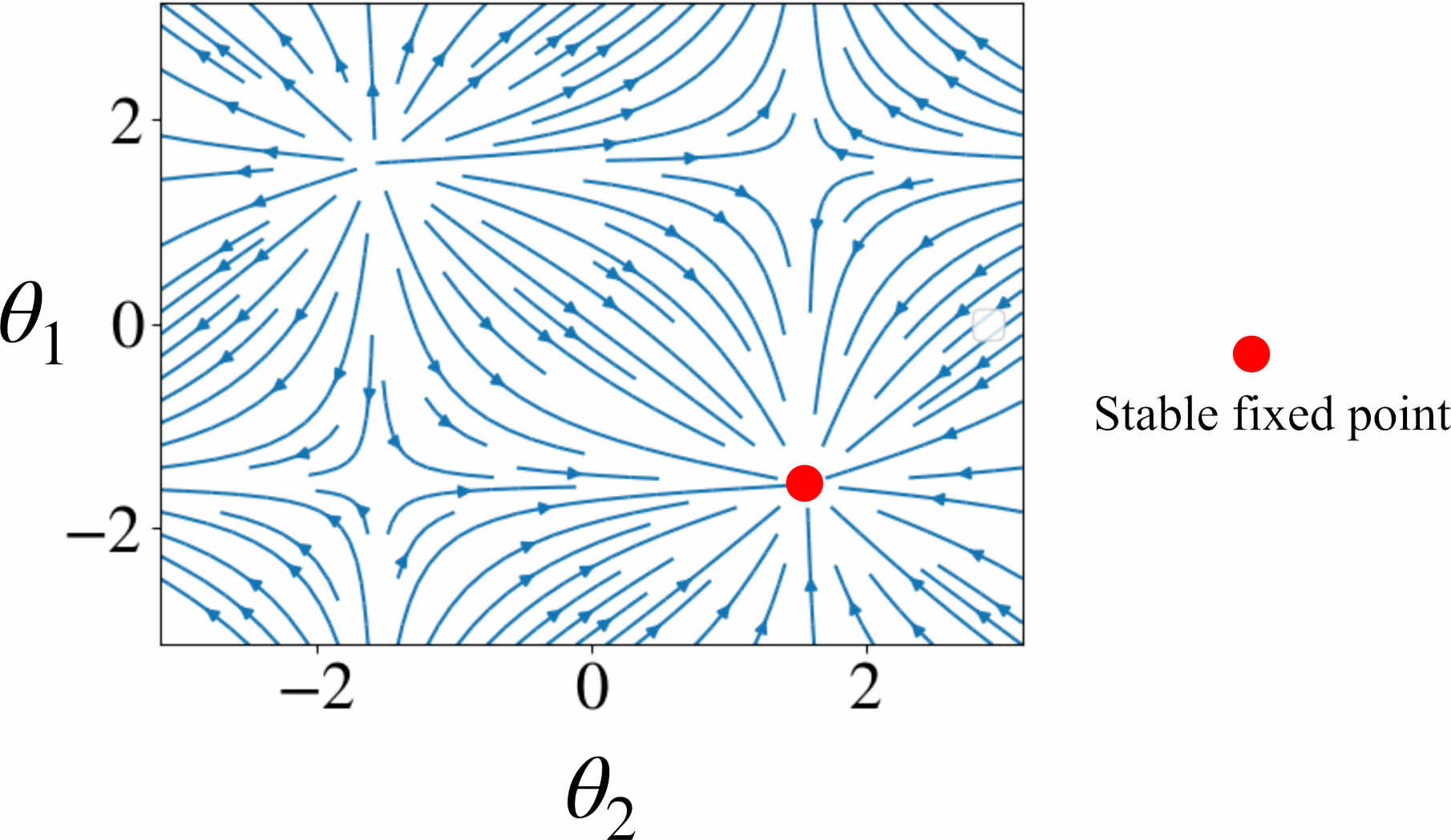}
    \caption{{Phase portrait of the dimer showing agreement with the linear stability analysis. See section \ref{sec:dimerAPP}}. }
    \label{fig:2pPhase}
\end{figure}

For the dynamics of a dimer (a chain with two monomers), Eqs.~\eqref{eq:pos_eom}--\eqref{eq:theta_eom} reduce from $\{x_i,y_i,\theta_i\}$ to the three coordinates $\{X_{\rm com},\theta_1,\theta_2\}$, governed by (ignoring noise)
\begin{subequations}
\begin{align}
\dot{X}_{\rm com} &= \frac{v_s}{2}\Big(\cos\theta_1+\cos\theta_2\Big), \\
\dot{\theta_1} &= -a\cos\theta_1, \\
\dot{\theta_2} &= +a\cos\theta_2,
\end{align}
\label{eq:dimerAPP_eom}
\end{subequations}

with $a\equiv\chi_r R=\chi_r\lambda_m/(2b)^2$ as in Eq.~\eqref{eq:dimer_eom}, using that in these reduced coordinates $x_1(t)-x_2(t)=0$ and $y_1(t)-y_2(t)=-2b$ for all $t$ (the rigid, fixed bond of IC0, Eq.~\eqref{eq:ic0}). Equations~\eqref{eq:dimerAPP_eom} are then linearized about the fixed point $(\theta_1,\theta_2)=(-\tfrac{\pi}{2},\tfrac{\pi}{2})$ -- equivalently $\theta_2=-\theta_1$, the mirror symmetry of Eq.~\eqref{eq:mirror_symmetry} -- to give
\begin{align}
\begin{pmatrix}
\delta\dot X_{\rm com} \\ \delta\dot\theta_1 \\ \delta\dot\theta_2
\end{pmatrix}
=
\begin{pmatrix}
0 & \tfrac{v_s}{2} & -\tfrac{v_s}{2} \\
0 & -a & 0 \\
0 & 0 & -a
\end{pmatrix}
\begin{pmatrix}
\delta X_{\rm com} \\ \delta\theta_1 \\ \delta\theta_2
\end{pmatrix},
\label{eq:lsa_mat_aabb}
\end{align}
with eigenvalues
\begin{align}
\lambda = 0, \quad -a, \quad -a,
\label{eq:eigs_aa}
\end{align}
the second twice degenerate. The marginal ($\lambda=0$) direction is $X_{\rm com}$, exactly as found for the single-angle reduction of Sec.~\ref{sec:dimers_det}; the two independent, twice-degenerate rates $-a=-\chi_r R$ recover the same eigenvalue quoted there, now obtained without first imposing $\theta_2=-\theta_1$ as an ansatz. In Fig.~\ref{fig:2pPhase}, we show that this agrees well with the numerical computation of the phase portrait.

\section{Equilibrium dynamics of the dipolar dimer}\label{app:dipole_dimer_eq}

\subsection{Reduction of the current and the equations of motion}

For $N=2$ the dipolar current of Eq.~\eqref{eq:dipolar_current} involves only the \emph{other} monomer's orientation, and using $\hat{\mathbf{r}}_{21}=-\hat{\mathbf{r}}_{12}\equiv-\hat r$,
\begin{align}
\mathbf{J}_1 = -\lambda_d\big[(\hat{\mathbf{e}}_2\cdot\hat r)\hat r-\hat{\mathbf{e}}_2/3\big]/r^3,\qquad
\mathbf{J}_2 = -\lambda_d\big[(\hat{\mathbf{e}}_1\cdot\hat r)\hat r-\hat{\mathbf{e}}_1/3\big]/r^3,
\end{align}
i.e. the same kernel with $1\leftrightarrow2$ (here $r=2b$). Computing $\dot\theta_i=\hat{\mathbf{e}}_i\times\mathbf{J}_i$ gives, with $\hat r=\hat{\mathbf{y}}$,
\begin{align}
\dot\theta_1 = -\frac{\lambda_d}{3r^3}\big[2\cos\theta_1\sin\theta_2+\sin\theta_1\cos\theta_2\big],\qquad
\dot\theta_2 = -\frac{\lambda_d}{3r^3}\big[2\cos\theta_2\sin\theta_1+\sin\theta_2\cos\theta_1\big],
\end{align}
a coupled pair, confirming that $\mathbf{J}_i$ depending on $\theta_j$ genuinely couples the two orientations (unlike the monopolar case). Defining $\Sigma=\theta_1+\theta_2$, $\Delta=\theta_1-\theta_2$, direct substitution gives, for \emph{arbitrary} $\theta_1,\theta_2$ (not merely near a fixed point),
\begin{align}
\dot\Sigma = \dot\theta_1+\dot\theta_2 = -\frac{\lambda_d}{r^3}\sin\Sigma, \qquad
\dot\Delta = \dot\theta_1-\dot\theta_2 = \frac{\lambda_d}{3r^3}\sin\Delta.
\end{align}
Each depends only on its own variable: the system decouples exactly, globally, into two independent one-dimensional gradient dynamicss,
\begin{align}
\dot\Sigma = -\frac{dV_\Sigma}{d\Sigma}, \quad V_\Sigma(\Sigma)=-b_\Sigma\cos\Sigma, \qquad
\dot\Delta = -\frac{dV_\Delta}{d\Delta}, \quad V_\Delta(\Delta)=b_\Delta\cos\Delta,
\end{align}
with $b_\Sigma=\lambda_d/r^3$, $b_\Delta=\lambda_d/(3r^3)\equiv a_{\rm dip}$ ($b_\Sigma=3a_{\rm dip}$). $V_\Sigma$ is minimized at $\Sigma^*=0$; $V_\Delta$ at $\Delta^*=\pi$. Equivalently, $(\theta_1,\theta_2)$ undergoes an exact two-body gradient dynamics of the joint potential $U(\theta_1,\theta_2)\propto\cos(\theta_1-\theta_2)-3\cos(\theta_1+\theta_2)$, for a suitable (positive) mobility constant. We verified $\dot\Sigma,\dot\Delta$ above against the torque computed by the full multi-particle engine at arbitrary (non-fixed-point, non-antisymmetric) $(\theta_1,\theta_2)$, finding close numerical agreement.

At the fixed point, $\Delta^*=\pi\Rightarrow\theta_1^*-\theta_2^*=\pi$ (propulsion directions exactly antiparallel), and $\Sigma^*=0\Rightarrow\theta_2^*=-\theta_1^*$, recovering the same antisymmetric structure as the monopolar dimer (Sec.~\ref{sec:dimers_det}), up to which of the two (degenerate, $\theta_1^*=\pm\pi/2$) branches is selected by symmetry-breaking in the initial condition. The center-of-mass velocity, with the internal spring force cancelling exactly in $\dot{\mathbf{X}}_{\rm com}$ as in Sec.~\ref{sec:dimers_det}, is
\begin{align}
\dot X_{\rm com} = v_s\cos(\Sigma/2)\cos(\Delta/2),
\label{eq:dipdimer_xcom}
\end{align}
which vanishes exactly at $\Delta^*=\pi$ ($\cos(\pi/2)=0$), independent of $\Sigma^*$: the halting mechanism is carried entirely by the $\Delta$ mode.

\subsection{Phase diagrams in $(\lambda_m,\lambda_d)$}\label{app:dipole_phase_diagram}

\begin{figure}[t]
\centering
\includegraphics[width=0.95\textwidth]{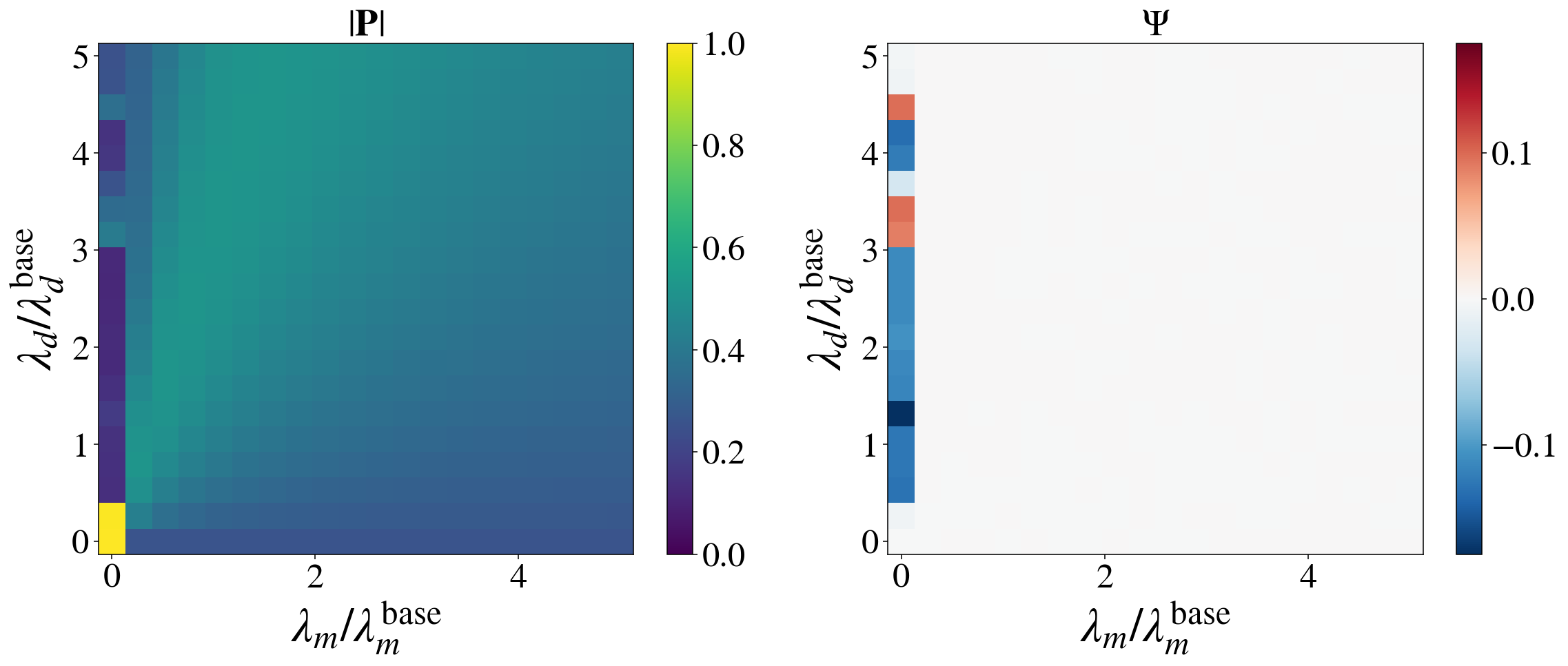}
\caption{Phase diagram in $(\lambda_m,\lambda_d)$, IC0, positive quadrant. Left: global polarization magnitude $|\mathbf{P}|$. Right: $\Psi=\mathbf{P}\cdot\hat{\mathbf{t}}$, on a diverging scale centered at zero. $\Psi\approx0$ across the whole $\lambda_m>0$ region confirms the exact C-shape perpendicularity condition; the nonzero, non-monotonic structure at $\lambda_m=0$ reflects the absence of any curling mechanism at that boundary.}
\label{fig:phase_diagram_ic0}
\end{figure}

To place the C-shape criterion of Eq.~\eqref{eq:psi_defn} in context across the full space of couplings, we computed both $|\mathbf{P}|$ and $\Psi$ at $t=T$ on a $20\times20$ grid spanning $\lambda_m,\lambda_d\in[0,5]\lambda^{\rm base}$ (in units of the respective base values used throughout Sec.~\ref{sec:dipole}) Fig.~\ref{fig:phase_diagram_ic0}. $\Psi$ is indistinguishable from zero across essentially the entire $\lambda_m>0$ half-plane, regardless of $\lambda_d$, confirming that the exact perpendicularity condition of Eq.~\eqref{eq:psi_defn} is reached robustly once monopolar coupling is present at all; $|\mathbf{P}|$ over the same region shows that this is not because $\mathbf P$ vanishes -- the chain is not depolarized -- but because $\mathbf P$ and the body axis remain exactly perpendicular as required.

\subsection{Stochastic solution: MSD and EPR}

Adding independent rotational noise $D_r$ to each $\theta_i$, the induced noise on $\Sigma,\Delta$ is itself independent between the two modes (their covariance vanishes, since $\mathrm{Var}(\theta_1)=\mathrm{Var}(\theta_2)$), each with effective strength $2D_r$. Both remain exact, independent one-dimensional Langevin equations, so the linearized (OUP) MSD of each mode is of the same form as Eq.~\eqref{eq:dimer_msd},
\begin{align}
\langle(\Delta\Sigma)^2\rangle(t) = \frac{4D_r}{b_\Sigma}\big(1-e^{-b_\Sigma t}\big), \qquad
\langle(\Delta\Delta)^2\rangle(t) = \frac{4D_r}{a_{\rm dip}}\big(1-e^{-a_{\rm dip} t}\big).
\end{align}
Since $\theta_1=(\Sigma+\Delta)/2$ and $\Sigma,\Delta$ are independent, the MSD of the physical angle itself is the \emph{sum} of two exponentials rather than one,
\begin{align}
\langle(\Delta\theta_1)^2\rangle(t) = D_r\left[\frac{1}{3a_{\rm dip}}\big(1-e^{-3a_{\rm dip} t}\big) + \frac{1}{a_{\rm dip}}\big(1-e^{-a_{\rm dip} t}\big)\right],
\label{eq:dipdimer_msd}
\end{align}
saturating at plateau $\frac{4}{3}(D_r/a_{\rm dip})$ -- two-thirds of what a single-mode formula with rate $a_{\rm dip}$ alone would give (Eq.~\eqref{eq:dimer_msd}) -- a genuine, quantifiable signature of the two-mode structure, though qualitatively the same OUP-trapped saturation as the monopolar dimer.

Since each of $\Sigma,\Delta$ is an exact 1D gradient dynamics with FDT-consistent (matching) noise, each satisfies detailed balance individually, so the total orientational-sector EPR vanishes identically, $\sigma_\theta\equiv0$ at any $D_r$ -- exactly as for the monopolar dimer (Sec.~\ref{sec:dimers_stoch}), now established via the sum of two independently-vanishing sectors rather than one. The exact stationary distribution of $\Delta$ is von Mises, $P_{ss}(\Delta)\propto\exp[k_\Delta\cos(\Delta-\pi)]$, $k_\Delta=a_{\rm dip}/(4D_r)$ (the factor of 4, rather than 2 as in Eq.~\eqref{eq:dimer_vonmises}, follows from the doubled effective noise strength $2D_r$). Using $\cos^2(\Delta/2)=(1+\cos\Delta)/2$ and the standard von Mises moment $\langle\cos(\Delta-\pi)\rangle_{ss}=I_1(k_\Delta)/I_0(k_\Delta)$, Eq.~\eqref{eq:dipdimer_xcom} gives the full system EPR in closed form,
\begin{align}
\sigma_{\rm dip} = \frac{v_s^2}{D_t}\left[1-\frac{I_1(k_\Delta)}{I_0(k_\Delta)}\right], \qquad k_\Delta=\frac{a_{\rm dip}}{4D_r},
\label{eq:dipdimer_epr}
\end{align}
which, like Eq.~\eqref{eq:dimer_epr}, vanishes as $D_r\to0$ ($k_\Delta\to\infty$, $I_1/I_0\to1$) and saturates at the same ceiling $v_s^2/D_t$ as $D_r\to\infty$ ($k_\Delta\to0$) -- the fully-randomized limit is insensitive to the coupling mechanism. We do not reproduce the corresponding MSD/histogram/EPR-vs-$D_r$ figures here: numerically, they are qualitatively identical in form to Fig.~\ref{fig:msd_hist}(a),(b),(d) (the monopolar dimer panels), consistent with Eqs.~\eqref{eq:dipdimer_msd}--\eqref{eq:dipdimer_epr} both reducing to the same qualitative OUP-trapped, detailed-balance structure.

\bibliography{refs_combined}

\begin{thebibliography}{27}
\bibitem [{Sup()}]{SupplementalMaterial}%
  \BibitemOpen
  \href@noop {} {\bibinfo {title} {See {S}upplemental {M}aterial at [url will be inserted by publisher] for videos of the stochastic and deterministic dynamics of the dimer (${N}=2$) and ${N}=6$ chain, under monopolar and dipolar coupling}}\BibitemShut {NoStop}%
\end{thebibliography}

\end{document}